\documentclass{cernyrep} 
\usepackage{texnames}
\usepackage[T1]{fontenc}
\usepackage{graphicx}
\usepackage{subfigure}
\usepackage{pstricks,pstricks-add, multido, pst-coil,pst-solides3d,pst-grad,pst-optic,pst-text}
\usepackage{slantsc}
\usepackage{psfrag}
\newcommand{\qed}{\nobreak \ifvmode \relax \else
      \ifdim\lastskip<1.5em \hskip-\lastskip
      \hskip1.5em plus0em minus0.5em \fi \nobreak
      \vrule height0.75em width0.5em depth0.25em\fi}
\newcommand{\mywidth}{0.7\linewidth}

\begin{document}
\title{RF measurements I: signal receiving techniques}
 
\author{F. Caspers}

\institute{CERN, Geneva, Switzerland}

\maketitle 

\begin{abstract}
For the characterization of components, systems and signals in the RF and microwave range, several dedicated instruments are in use. In this paper the fundamentals of the RF-signal sampling technique, which has found widespread applications in `digital' oscilloscopes and sampling scopes, are discussed. The key element in these front-ends is the Schottky diode which can be used either as an RF mixer or as a single sampler. The spectrum analyser has become an absolutely indispensable tool for RF signal analysis. Here the front-end is the RF mixer as the RF section of modern spectrum analysers has a rather complex architecture. The reasons for this complexity and certain working principles as well as limitations are discussed. In addition, an overview of the development of scalar and vector signal analysers is given. For the determination of the noise temperature of a one-port and the noise figure of a two-port, basic concepts and relations are shown. A brief discussion of commonly used noise measurement techniques concludes the paper.
\end{abstract}
 
\section{Introduction}
In the early days of RF engineering the available instrumentation for measurements was rather limited. Besides elements acting on the heat developed by RF power (bimetal contacts and resistors with very high temperature coefficient) only point/contact diodes, and to some extent vacuum tubes, were available as signal detectors. For several decades the slotted measurement line \cite{smith} was the most used instrument for measuring impedances and complex reflection coefficients. Around 1960 the tedious work with such coaxial and waveguide measurement lines became considerably simplified with the availability of the vector network analyser. At the same time the first sampling oscilloscopes with 1\,GHz bandwidth arrived on the market. This was possible due to progress in solid-state (semiconductor) technology and advances in microwave elements (microstrip lines). Reliable, stable, and easily controllable microwave sources are the backbone of spectrum and network analysers as well as sensitive (low noise) receivers. This paper will only treat signal receiving devices such as spectrum analysers and oscilloscopes. For an overview of network analysis tools see \textit{RF measurements II: network analysis}. 
\section{Basic elements and concepts}
Before discussing several measurement devices, a brief overview of the most important components in such devices and some basic concepts are presented. 
\subsection{Decibel}
Since the unit dB is frequently used in RF engineering a short introduction and definition of terms is given here. The decibel is the unit used to express relative differences in signal power. It is expressed as the base 10 logarithm of the ratio of the powers of two signals: 
\begin{equation}
 P\text{ [dB]} = 10 \cdot \text{log}(P/P_{0})\hspace{0.2cm}.
\label{dbpower}
\end{equation}
It is also common to express the signal amplitude in dB. Since power is proportional to the square of a signal's amplitude, the voltage in dB is expressed as follows: 
\begin{equation}
V\text{ [dB]} = 20 \cdot \text{log}(V/V_{0})\hspace{0.2cm}.
\label{dbvoltage}
\end{equation}
In Eqs.\,(\ref{dbpower}) and (\ref{dbvoltage}) $P_{0}$ and $V_{0}$ are the reference power and voltage, respectively. A given value in dB is the same for power ratios as for voltage ratios. Please note that there are no `power dB' or `voltage dB' as dB values always express a ratio.

Conversely, the absolute power and voltage can be obtained from dB values by
\begin{eqnarray}
P = P_{0} \cdot 10^{\frac{P\text{ [dB]}}{10}}\hspace{0.2cm}, \\
V = V_{0} \cdot 10^{\frac{V\text{ [dB]}}{20}}\hspace{0.2cm}.
\label{conversion}
\end{eqnarray}

Logarithms are useful as the unit of measurement because
\begin{enumerate}
	\item signal power tends to span several orders of magnitude and
	\item signal attenuation losses and gains can be expressed in terms of subtraction and addition.
\end{enumerate}
Table \ref{dB} helps to indicate the order of magnitude associated with dB.
\begin{table}
\centering
\caption{Overview of dB key values and their conversion into power and voltage ratios.}
\begin{tabular}{ccc}
\hline
\hline
 & Power ratio & Voltage ratio \\
\hline
$-$20 dB & 0.01 & 0.1 \\
$-$10 dB & 0.1 & 0.32 \\
$-$3 dB & 0.50 & 0.71 \\
$-$1 dB & 0.74 & 0.89 \\
0 dB & 1 & 1 \\
1 dB & 1.26 & 1.12 \\
3 dB & 2.00 & 1.41 \\
10 dB & 10 & 3.16 \\
20 dB & 100 & 10 \\
$n \cdot $10 dB & 10$^{n}$ & 10$^{n\text{/2}}$ \\
\hline
\hline
\end{tabular}
\label{dB}
\end{table}

Frequently dB values are expressed using a special reference level and not SI units. Strictly speaking, the reference value should be included in parentheses when giving a dB value, e.g., +3 dB (1\,W) indicates 3 dB at $P_{0}$ = 1\,watt, thus 2\,W. However, it is more common to add some typical reference values as letters after the unit, for instance, dBm defines dB using a reference level of $P_{0}$ = 1 mW. 
Thus, 0 dBm correspond to $-$30 dBW, where dBW indicates a reference level of $P_{0}$ = 1\,W. Often a reference impedance of 50\,$\Omega$ is assumed. 
Other common units are
\begin{itemize}
	\item dBmV for the small voltages with $V_{0}$ = 1\,mV and
	\item dBmV/m for the electric field strength radiated from an antenna with reference field strength $E_{0}$ = 1 mV/m
\end{itemize}
\subsection{The RF diode}
One of the most important elements inside all sophisticated measurement devices is the fast RF diode or Schottky diode. The basic metal--semiconductor junction has an intrinsically very fast switching time of well below a picosecond, provided that the geometric size and hence the junction capacitance of the diode is small enough. However, this unavoidable and voltage dependent junction capacity will lead to limitations of the maximum operating frequency. 

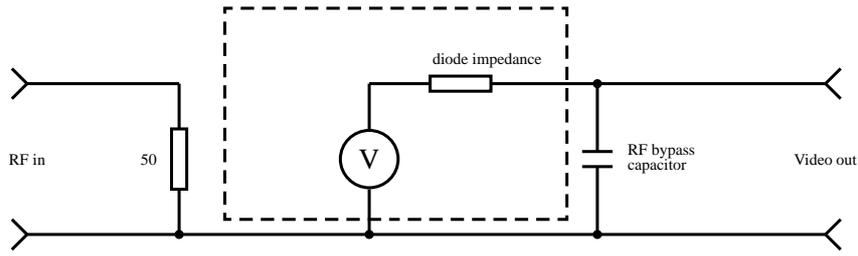
\begin{figure}[htbp]%
\centering
%
\begin{pspicture}(0,0)(12.5,5)%
\psset{linewidth=0.04}%
\psline(1,1)(11.5,1)%
\psline(0.8,0.8)(1,1)%
\psline(0.8,1.2)(1,1)%
\psline(11.7,0.8)(11.5,1)%
\psline(11.7,1.2)(11.5,1)%
\psline(1,3)(3,3)%
\psline(0.8,2.8)(1,3)%
\psline(0.8,3.2)(1,3)%
\psline(3,3)(3,2.4)%
\pspolygon(2.9,2.4)(3.1,2.4)(3.1,1.6)(2.9,1.6)%
\psline(3,1)(3,1.6)%
\psline(5.5,1)(5.5,1.6)%
\pscircle(5.5,2){0.4}%
\rput(5.5,2){V}%
\psline(5.5,2.4)(5.5,3)%
\psline(5.5,3)(6.3,3)%
\pspolygon(6.3,2.9)(7.1,2.9)(7.1,3.1)(6.3,3.1)%
\psline(7.1,3)(8.5,3)%
\psline(8.5,3)(8.5,2.1)%
\psline(8.3,2.1)(8.7,2.1)%
\psline(8.3,1.9)(8.7,1.9)%
\psline(8.5,1.9)(8.5,1)%
\psline(8.5,3)(11.5,3)%
\psline(11.7,2.8)(11.5,3)%
\psline(11.7,3.2)(11.5,3)%
\pspolygon[linestyle=dashed](3.6,1.2)(3.6,4)(8.1,4)(8.1,1.2)%
\rput(3,1){%
\pscircle[fillstyle=solid,fillcolor=black](0,0){0.06}%
}%
\rput(5.5,1){%
\pscircle[fillstyle=solid,fillcolor=black](0,0){0.06}%
}%
\rput(8.5,1){%
\pscircle[fillstyle=solid,fillcolor=black](0,0){0.06}%
}%
\rput(8.5,3){%
\pscircle[fillstyle=solid,fillcolor=black](0,0){0.06}%
}%
\rput(1,2){\tiny RF in}%
\rput(11.5,2){\tiny Video out}%
\rput[l](8.9,2.1){\tiny RF bypass}%
\rput[l](8.9,1.9){\tiny capacitor}%
\rput(7.05,3.3){\tiny diode impedance}%
\rput[r](2.7,2){\tiny 50}%
\end{pspicture}
%
%
\caption{The equivalent circuit of a diode}%
\label{circdiode}%
\end{figure}

\begin{figure}[htbp]%
\centering
\includegraphics[width=\mywidth]{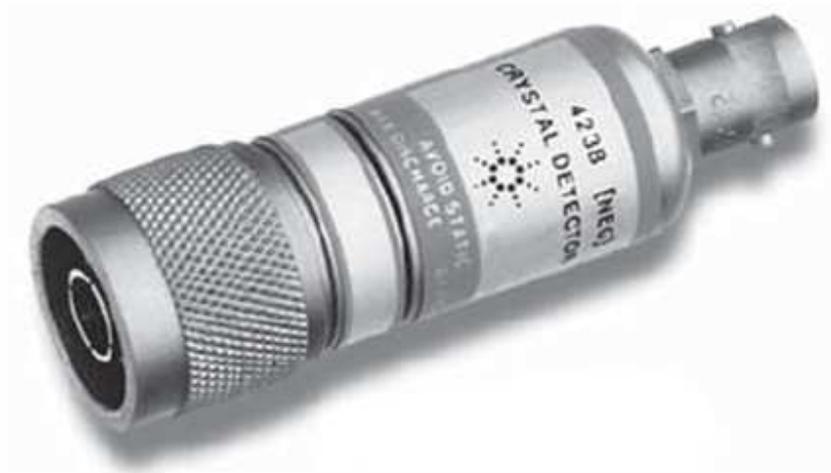}%
\caption{A commonly used Schottky diode. The RF input of this detector diode is on the left and the video output on the right (figure courtesy Agilent)}%
\label{diode1}%
\end{figure}

The equivalent circuit of such a diode is depicted in Fig.\,\ref{circdiode} and
an example of a commonly used Schottky diode can be seen in Fig.\,\ref{diode1}.
One of the most important properties of any diode is its characteristic which is the relation of current as a function of voltage. This relation is described by the Richardson equation \cite{vendelin}:
\begin{equation}
I = AA_{\text{RC}}T^{2}\text{exp}\left( - \frac{q\phi_{\text{B}}}{kT}\right)\left[\text{exp}\left(\frac{qV_{\text{J}}}{kT}\right) - \text{M}\right]\hspace{0.2cm},
\label{richard}
\end{equation}
where $A$ is the area in cm$^{2}$, $A_{\text{RC}}$ the modified Richardson constant, $k$ Boltzmann's constant, $T$ the absolute temperature, $\phi_{\text{B}}$ the barrier height in volts, $V_{\text{J}}$ the external Voltage across the depletion layer, M the avalanche multiplication factor and $I$ the diode current.

\begin{figure}[htbp]%
\centering
\input{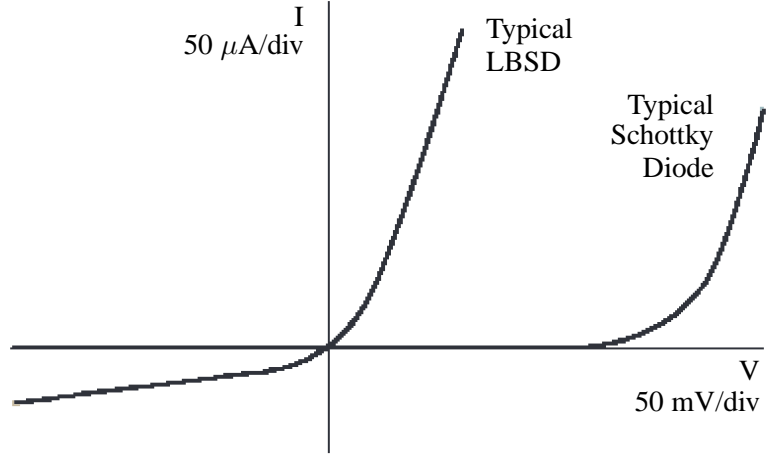}
\caption{Current as a function of voltage for different diode types (LBSD = low barrier Schottky diode)}%
\label{kenn}%
\end{figure}

\begin{figure}[htbp]%
\centering
\begin{pspicture}(0,0)(6,4)%
\psset{linewidth=0.04}%
\psline[arrows=->](1,0.5)(5,0.5)%
\psline[arrows=->](1,0.5)(1,4)%
\psline[linecolor=red](1,0.5)(3,0.5)%
\psline[linecolor=red](3,0.5)(3,3)%
\psline[linecolor=red,linestyle=dotted](3,3)(3,4)%
\psline[arrows=->](4.1,1.6)(3.1,0.6)%
\rput[l](4.1,1.7){Threshold voltage }%
\rput(5,0){Voltage}%
\rput[l](1.2,3.8){Current}
\end{pspicture}%
\caption{The current--voltage relation of an ideal commutator with threshold voltage}%
\label{comm}%
\end{figure}
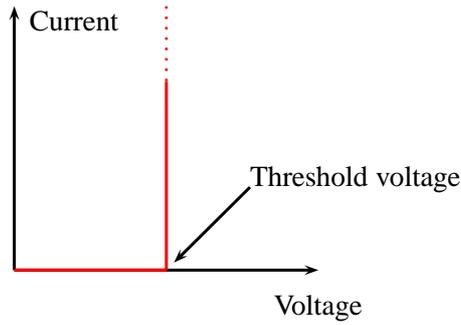

This relation is depicted graphically for two diodes in Fig.\,\ref{kenn}.
As can be seen, the diode is not an ideal commutator (Fig.\,\ref{comm}) for small signals. Note that it is not possible to apply big signals, since this kind of diode would burn out.
However, there exist rather large power versions of Schottky diodes which can stand more than 9\,kV and several 10\,A but they are not suitable in microwave applications due to their large junction capacity.

The Richardson equation can be roughly approximated by a simpler equation \cite{vendelin}:
\begin{equation}
I = I_{\text{s}}\left[\text{exp} \left(\frac{V_{\text{J}}}{0.028}\right) - 1\right]\hspace{0.2cm}.
\label{richsimple}
\end{equation}
This approximation can be used to show that the RF rectification is linked to the second derivation (curvature) of the diode characteristic.

\begin{figure}[htbp]%
\centering
\begin{pspicture}(0,0)(15,9.5)%
\newrgbcolor{olive}{0.6 0.6 0}%
\newrgbcolor{cerulean}{0.1 0.6 0.9}%
\newrgbcolor{ocker}{0.7 0.4 0}%
\psset{linewidth=0.01}%
\multirput(0,0)(2.5,0){6}{\psline(1,1)(1,8.5)}%
\multirput(0,0)(0,1.5){6}{\psline(1,1)(13.5,1)}%
\psset{linewidth=0.04}%
\psline[linestyle=dashed,linecolor=cerulean](1,1.4)(12.1,8.5)%
\psline[linestyle=dashed,linecolor=ocker](2.4,1)(13.5,8.1)%
\psline[linecolor=olive](1,1.4)(9,6.518)%
\psline[linecolor=olive](2.4,1)(9,5.2216)%
\psplot[algebraic=true,linecolor=olive]{9}{11}{-0.15*x*x+3.25*x-10.6}%
\rput(0,-1.2964){\psplot[algebraic=true,linecolor=olive]{9}{11}{-0.15*x*x+3.25*x-10.6}}%
\rput[r](4,3.6){\cerulean without load}%
\rput[l](7,3.6){\ocker square law loaded}%
\rput[l](9.6,5){\olive LBSD}%
\rput[r](8,6.2){\olive LBSD}%
\rput(13.5,0.7){0}%
\rput(11,0.7){-10}%
\rput(8.5,0.7){-20}%
\rput(6,0.7){-30}%
\rput(3.5,0.7){-40}%
\rput(1,0.7){-50}%
\rput[r](0.8,1){0.005}%
\rput[r](0.8,2.5){0.05}%
\rput[r](0.8,4){0.5}%
\rput[r](0.8,5.5){5.0}%
\rput[r](0.8,7){50}%
\rput[r](0.8,8.5){500}%
\rput(7.25,0.2){Input power [dBm]}%
\rput{90}(0,4.75){Output power [mV]}%
\end{pspicture}%
\caption{Relation between input power and output voltage}%
\label{squarelaw}%
\end{figure}
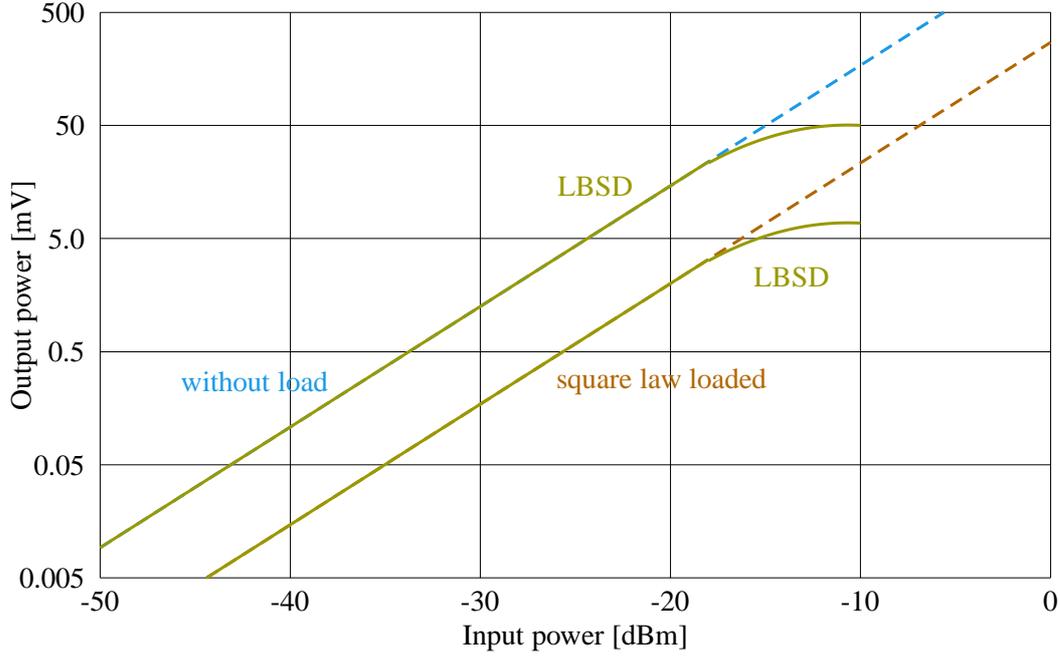

If the DC current is held constant by a current regulator or a large resistor assuming external DC bias\footnote{Most diodes do not need an external bias, since they have a DC return self-bias.}, then the total junction current, including RF is
\begin{equation}
I = I_{0} = \text{i}_{0} \text{\,cos\,} \omega t
\label{skew1}
\end{equation}
and hence the current--voltage relation can be written as
\begin{equation}
V_{\text{J}} = 0.028 \text{ ln}\left(\frac{I_{\text{S}} + I_{0} + \text{i cos\,} \omega t}{I_{\text{S}}}\right) = 0.028 \text{ ln}\left(\frac{I_{0} + I_{\text{S}}}{I_{\text{S}}}\right) + 0.028 \text{ ln}\left(\frac{\text{i cos\,}\omega t}{I_{0} + I_{\text{S}}}\right)\hspace{0.2cm}.
\label{skew2}
\end{equation}
If the RF current $I$ is small enough, the second term can be approximated by Taylor expansion: 
\begin{equation}
V_{\text{J}} \approx 0.028 \text{ ln}\left(\frac{I_{0} + I_{\text{S}}}{I_{\text{S}}}\right) + 0.028\left[\frac{\text{i cos\,}\omega t}{I_{0} + I_{\text{S}}} - \frac{\text{i}^{2} \text{cos}^{2} \omega t}{2(I_{0} + I_{\text{S}})^{2}} + \ldots\right] = V_{\text{DC}} + V_{\text{J}}\text{\,cos\,} \omega t + \text{higher order terms}
\label{skew3}
\end{equation}
With the identity cos$^{2} = 0.5$, the DC and the RF voltages are given by
\begin{equation}
V_{\text{J}} = \frac{0.028}{I_{0} + I_{\text{S}}} \text{i} = R_{\text{S}} \text{i\hspace{0.2cm}and\hspace{0.2cm}} V_{\text{DC}} = 0.028 \text{ ln}\left(1 + \frac{I_{0}}{I_{\text{S}}}\right) - \frac{0.028^{2}}{4(I_{0} + I_{\text{S}})^{2}} = V_{0} - \frac{V_{\text{J}}^{2}}{0.112}\hspace{0.2cm}.
\label{skew4}
\end{equation}
The region where the output voltage is proportional to the input power is called the square-law region (Fig.\,\ref{squarelaw}).
In this region the input power is proportional to the square of the input voltage and the output signal is proportional to the input power, hence the name square-law region. 

The transition between the linear region and the square-law region is typically between $-$10 and $-$20\,dB (Fig.\,\ref{squarelaw}).

There are fundamental limitations when using diodes as detectors. The output signal of a diode (essentially DC or modulated DC if the RF is amplitude modulated) does not contain a phase information. In addition, the sensitivity of a diode restricts the input level range to about $-$60\,dBm at best which is not sufficient for many applications. 

The minimum detectable power level of an RF diode is specified by the `tangential sensitivity' which typically amounts to $-$50 to $-$55\,dBm for 10\,MHz video bandwidth at the detector output \cite{thumm}.

To avoid these limitations, another method of operating such diodes is needed.
\subsection{Mixer}
\begin{figure}[htbp]%
\centering
\includegraphics[width=\mywidth]{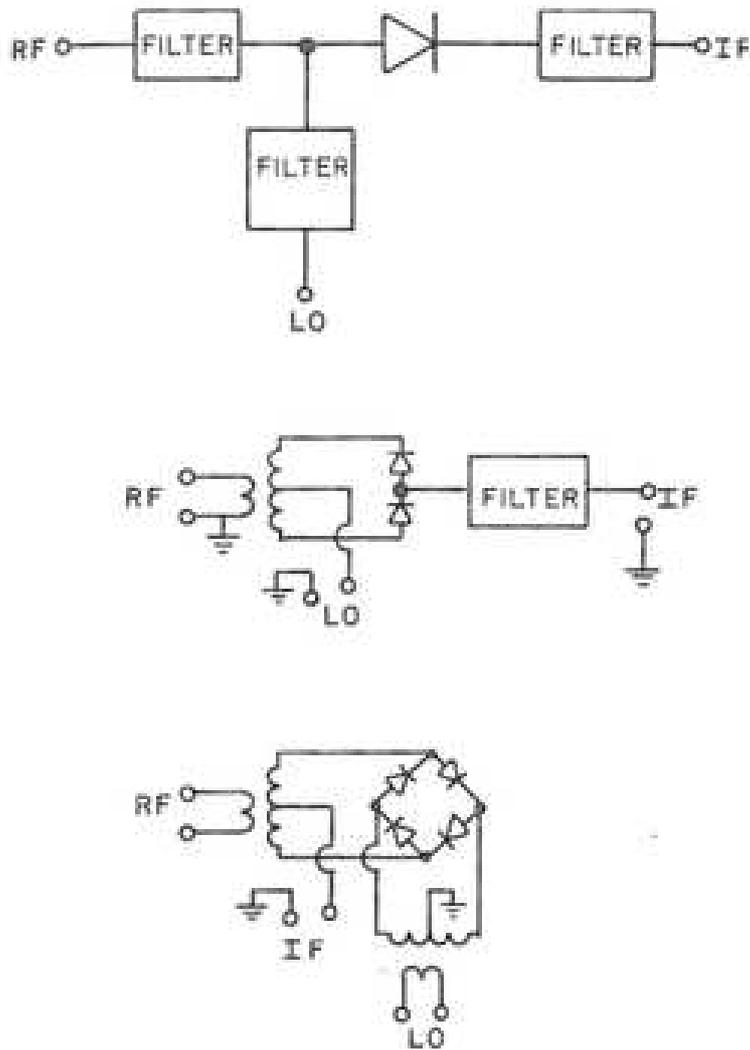}%
\caption{Examples of different mixer configurations}%
\label{mixers}%
\end{figure}

For the detection of very small RF signals a device that has a linear response over the full range (from 0 dBm ( = 1mW) down to thermal noise = $-$174\,dBm/Hz = 4$\cdot$10$^{-21}$\,W/Hz) is preferred. An RF mixer provides these features using 1, 2, or 4 diodes in different configurations (Fig.\,\ref{mixers}). 
A mixer is essentially a multiplier with a very high dynamic range implementing the function
\begin{equation}
f_{1}(t) · f_{2}(t) \text{\hspace{0.2cm}with } f_{1}(t) = \text{ RF signal\hspace{0.2cm} and } f_{2}(t) = \text{ LO signal}\hspace{0.2cm},
\label{mixer1}
\end{equation}
or more explicitly for two signals with amplitude $a_{i}$ and frequency $f_{i}$ ($i = 1, 2$):
\begin{equation}
a_{1}\text{\,cos}(2\pi f_{1}t + \varphi) \cdot a_{2}\text{\,cos}(2\pi f_{2}t) = \frac{1}{2}a_{1}a_{2}\left[\text{cos}((f_{1} + f_{2})t + \varphi) + \text{cos}((f_{1} - f_{2})t + \varphi)\right]\hspace{0.2cm}.
\label{mixer2}
\end{equation}
Thus we obtain a response at the IF (intermediate frequency) port that is at the sum and difference frequency of the LO (local oscillator $ = f_{1}$) and RF ($ = f_{2}$) signals.

Examples of different mixer configurations are shown in Fig.\,\ref{mixers}.

\begin{figure}[htbp]%
\centering
\begin{pspicture}(0,0)(15,4)%
\psset{linewidth=0.04}%
\rput(-1,0){%
\psline(1,1)(1,1.5)
\pscircle(1,2){0.5}
\rput(0.7,2){\psplot[algebraic=true]{0}{0.6}{0.1*sin(10.4719755*x)}}
\psline(1,2.5)(1,3)
\psline(1,1)(3,1)
\psline(1,3)(3,3)
\pscircle(3.1,3){0.1}
\pscircle(3.1,1){0.1}
\psline[arrows=->](5,1)(3.2,1)%
\psline[arrows=->](5,3)(3.2,3)%
\pscircle(5.1,3){0.1}
\pscircle(5.1,1){0.1}
\psline(5.2,1)(7,1)%
\psline(5.2,3)(7,3)%
\psline(7,3)(7,2.4)%
\pspolygon(6.9,2.4)(7.1,2.4)(7.1,1.6)(6.9,1.6)%
\psline(7,1.6)(7,1)%
\psline[arrows=<->,linewidth=0.1](8,2)(9,2)%
\rput(8.5,2.4){LO}%
\rput(9,0){%
\psline(1,1)(1,1.5)
\pscircle(1,2){0.5}
\rput(0.7,2){\psplot[algebraic=true]{0}{0.6}{0.1*sin(10.4719755*x)}}
\psline(1,2.5)(1,3)
\psline(1,1)(3,1)
\psline(1,3)(3,3)
\pscircle(3.1,3){0.1}
\pscircle(3.1,1){0.1}
\psline[arrows=->](5,1)(3.2,3)%
\psline[arrows=->](5,3)(3.2,1)%
\pscircle(5.1,3){0.1}
\pscircle(5.1,1){0.1}
\psline(5.2,1)(7,1)%
\psline(5.2,3)(7,3)%
\psline(7,3)(7,2.4)%
\pspolygon(6.9,2.4)(7.1,2.4)(7.1,1.6)(6.9,1.6)%
\psline(7,1.6)(7,1)%
}%
}%
\end{pspicture}%
\caption{Two circuit configurations interchanging with the frequency of the LO where the switches represent the diodes}%
\label{mixprinc}%
\end{figure}
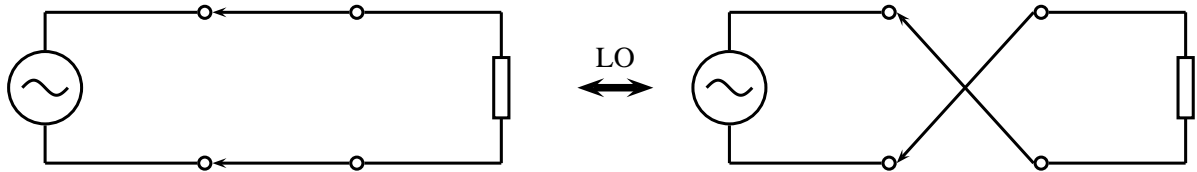

As can be seen from Fig.\,\ref{mixers}, the mixer uses diodes to multiply the two ingoing signals. These diodes function as a switch, opening different circuits with the frequency of the LO signal (Fig.\,\ref{mixprinc}).

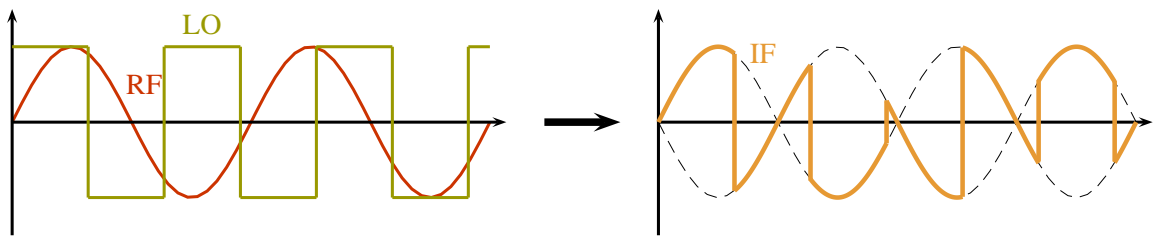
\begin{figure}[htbp]%
\centering
\begin{pspicture}(0,0)(15,5)%
\psset{linewidth=0.04}%
\newrgbcolor{myred}{0.8 0.2 0}%
\newrgbcolor{olive}{0.6 0.6 0}%
\newrgbcolor{orange}{0.9 0.6 0.2}%
\rput(0,-6){%
\rput(1,8){\psplot[algebraic=true,linecolor=myred]{0}{6.283185}{sin(2*x)}}%
\psline[arrows=->](1,8)(7.5,8)%
\psline[arrows=->](1,6.5)(1,9.5)%
\multirput(0,0)(2,0){3}{%
\psline[linecolor=olive](1,9)(2,9)%
}%
\multirput(0,0)(1,0){6}{%
\psline[linecolor=olive](2,9)(2,7)%
}%
\multirput(0,0)(2,0){3}{%
\psline[linecolor=olive](2,7)(3,7)%
}%
\psline[linecolor=olive](7,9)(7.283185,9)%
\rput[l](2.5,8.5){\myred RF}%
\rput(3.5,9.3){\olive LO}
\psline[arrows=->,linewidth=0.1](8,8)(9,8)%
\rput(8.5,0){%
\psline[arrows=->](1,8)(7.5,8)%
\psline[arrows=->](1,6.5)(1,9.5)%
\rput(1,8){\psplot[algebraic=true,linewidth=0.01,linestyle=dashed]{0}{6.283185}{sin(2*x)}}%
\rput(1,8){\psplot[algebraic=true,linewidth=0.01,linestyle=dashed]{0}{6.283185}{sin(2*x + 3.1415926)}}%
\psset{linewidth=0.06}
\rput(1,8){\psplot[algebraic=true,linecolor=orange]{0}{1}{sin(2*x)}}%
\rput(1,8){\psplot[algebraic=true,linecolor=orange]{2}{3}{sin(2*x)}}%
\rput(1,8){\psplot[algebraic=true,linecolor=orange]{4}{5}{sin(2*x)}}%
\rput(1,8){\psplot[algebraic=true,linecolor=orange]{6}{6.283185}{sin(2*x)}}%
\rput(1,8){\psplot[algebraic=true,linecolor=orange]{1}{2}{sin(2*x + 3.1415926)}}%
\rput(1,8){\psplot[algebraic=true,linecolor=orange]{3}{4}{sin(2*x + 3.1415926)}}%
\rput(1,8){\psplot[algebraic=true,linecolor=orange]{5}{6}{sin(2*x + 3.1415926)}}%
\psline[linecolor=orange](2,8.9051)(2,7.08961)%
\psline[linecolor=orange](3,8.75)(3,7.22432)%
\psline[linecolor=orange](4,8.289)(4,7.719)%
\psline[linecolor=orange](5,8.99)(5,7.01)%
\psline[linecolor=orange](6,8.56)(6,7.48)%
\psline[linecolor=orange](7,8.55)(7,7.48)%
\rput[l](2.2,8.9051){\orange IF}%
}%
}%
\end{pspicture}%
\caption{Time 
domain response of a mixer}%
\label{mixerresponse}%
\end{figure}

The response of a mixer in time 
domain is depicted in Fig.\,\ref{mixerresponse}.

The output signal is always in the `linear range' provided that the mixer is not in saturation with respect to the RF input signal. Note that for the LO signal the mixer should always be in saturation to make sure that the diodes work as a nearly ideal switch. 
The phase of the RF signal is conserved in the output signal available form the RF output.
\subsection{Amplifier}
\begin{figure}[htbp]%
\centering
\begin{pspicture}(0,0)(7,3)%
\psset{linewidth=0.04}%
\psline(0.2,1)(1,1)%
\pspolygon(1,0.4)(1,1.6)(2,1)%
\psline(2,1)(2.8,1)%
\rput(0.2,0.7){1}%
\rput(2.8,0.7){2}%
\rput[l](4,1){S = $\left(
\begin{array}{cc}
	0 & 0 \\
	G & 0
\end{array}
\right)$}%
\end{pspicture}%
%
\caption{Circuit symbol an S-matrix of an ideal amplifier}%
\label{ampli}%
\end{figure}
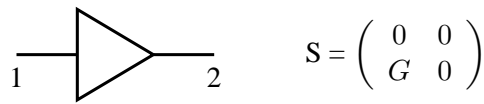

A linear amplifier augments the input signal by a factor which is usually indicated in decibel.
The ratio between the output and the input signal is called the transfer function and its magnitude---the voltage gain $G$---is measured in dB and given as
\begin{equation}
G [\text{dB}] = 20 \cdot \frac{\text{V}_{\text{RFout}}}{\text{V}_{\text{RFin}}} \text{\hspace{0.2cm}or\hspace{0.2cm} } \frac{\text{V}_{\text{RFout}}}{\text{V}_{\text{RFin}}} = 20 \cdot \text{log} G [\text{lin}]\hspace{0.2cm}.
\label{gain}
\end{equation}
The circuit symbol of an amplifier is shown in Fig.\,\ref{ampli} together with its S-matrix.

The bandwidth of an amplifier specifies the frequency range where it is usually operated. This frequency range is defined by the $-$3\,dB points\footnote{The $-$3\,dB points are the points left and right of a reference value (e.g., a local maximum of a curve) that are 3\,dB lower than the reference.} with respect to its maximum or nominal transmission gain.
%
%

\begin{figure}[htbp]%
\centering
\includegraphics[width=\mywidth]{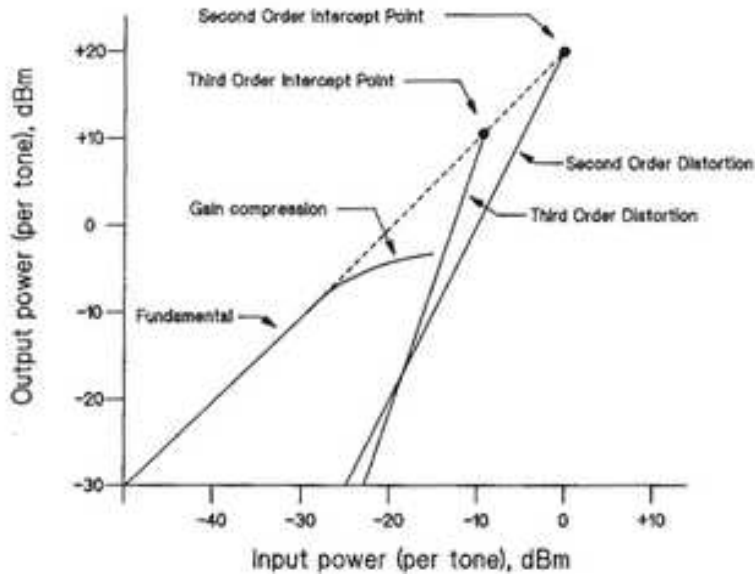}%
\caption{Example for the 1dB compression point \cite{witte}}%
\label{1dB}
\end{figure}

In an ideal amplifier the output signal would be proportional to the input signal. However, a real amplifier is nonlinear, such that for larger signals the transfer characteristic deviates from its linear properties valid for small signal amplification. When increasing the output power of an amplifier, a point is reached where the small signal gain becomes reduced by 1\,dB (Fig.\,\ref{1dB}). 
This output power level defines the 1\,dB compression point, which is an important measure of quality for any amplifier (low level as well as high power).

The transfer characteristic of an amplifier can be described in terms which are commonly used for RF engineering, i.e., the S-matrix (for further details see the paper on S-matrices of this School).
As implicitly contained in the S-matrix, the amplitude and phase information of any spectral component are preserved when passing through an ideal amplifier. For a real amplifier the element $G = \text{S}_{21}$ (transmission from port 1 to port 2) is not a constant but a complex function of frequency. Also the elements S$_{11}$ and S$_{22}$ are not 0 in reality.
\subsection{Interception points of nonlinear devices}
Important characteristics of nonlinear devices are the interception points. Here only a brief overview will be given. For further information the reader is referred to Ref.\,\cite{witte}.

One of the most relevant interception points is the interception point of 3rd order (IP3 point). Its importance derives from its straightforward determination, plotting the input versus the output power in logarithmic scale (Fig.\,\ref{1dB}). The IP3 point is usually not measured directly, but	extrapolated from measurement	data at much smaller power levels in order to avoid overload and damage of the device under test (DUT). If two signals $(f_{1},f_{2} > f_{1})$ which are	closely spaced by $\Delta f$ in frequency are	simultaneously applied to the DUT, the intermodulation products appear at	+ $\Delta f$ above $f_{2}$ and at $-$ $\Delta f$ below $f_{1}$. 


The transfer functions or weakly nonlinear devices can be approximated by Taylor expansion. Using $n$ higher order terms on one hand and plotting them together with an ideal linear device in logarithmic scale leads to two lines with different slopes ($x^{n} \stackrel{\text{log}}{\rightarrow} n \cdot \text{log }x$). Their intersection point is the intercept point of $n$th order. These points provide important information concerning the quality of nonlinear devices.

In this context, the aforementioned 1\,dB compression point of an amplifier is the intercept point of first order.

Similar characterization techniques can also be applied with mixers which with respect to the LO signal cannot be considered a weakly nonlinear device.

\subsection{The superheterodyne concept}
\begin{figure}[htbp]%
\centering
\begin{pspicture}(0,0)(15,7)%
\psset{linewidth=0.04}%
\newrgbcolor{yellow}{0.9 0.8 0.2}%
\newrgbcolor{myred}{0.8 0.2 0}%
\newrgbcolor{olive}{0.6 0.6 0}%
\newrgbcolor{orange}{0.9 0.6 0.2}%
\newrgbcolor{petrol}{0.2 0.5 0.6}%
\newrgbcolor{cerulean}{0.1 0.6 0.9}%
\newrgbcolor{grapefruit}{0.8 0.5 0.4}%
\newrgbcolor{ocker}{0.7 0.4 0}%
\psline(0.8,4)(0.8,6.6)%
\psline(0.4,6.6)(0.8,6)%
\psline(1.2,6.6)(0.8,6)%
\psline[arrows=->](0.8,4)(1.8,4)%
\pspolygon[fillcolor=yellow,fillstyle=solid](1.8,3.4)(1.8,4.6)(2.8,4)%
\psline[arrows=->](2.8,4)(3.8,4)%
\pspolygon[fillcolor=myred,fillstyle=solid](3.8,3.4)(5,3.4)(5,4.6)(3.8,4.6)%
\psline(3.8,3.4)(5,4.6)%
\psline(3.8,4.6)(5,3.4)%
\psline[arrows=->](4.4,1.2)(4.4,3.4)%
\pspolygon[fillcolor=olive,fillstyle=solid](3.8,1)(5,1)(5,2.2)(3.8,2.2)%
\psline(3.9,1.6)(4.1,1.6)%
\psline(4.2,1.1)(4.2,2.1)%
\pspolygon(4.3,1.2)(4.5,1.2)(4.5,2)(4.3,2)%
\psline(4.6,1.1)(4.6,2.1)%
\psline(4.6,1.6)(4.9,1.6)%
\psline[arrows=->](5,4)(6,4)%
\pspolygon[fillcolor=orange,fillstyle=solid](6,3.4)(7.2,3.4)(7.2,4.6)(6,4.6)%
\rput(6.6,4){BP}%
\psline[arrows=->](7.2,4)(8.2,4)%
\pspolygon[fillcolor=petrol,fillstyle=solid](8.2,3.4)(8.2,4.6)(9.2,4)%
\psline[arrows=->](9.2,4)(10.2,4)%
\pspolygon[fillcolor=cerulean,fillstyle=solid](10.2,3.4)(11.4,3.4)(11.4,4.6)(10.2,4.6)%
\psline(10.4,4)(11.2,4)%
\pspolygon(10.5,4)(10.5,4.3)(11.1,4)%
\psline(11.1,3.7)(11.1,4.3)%
\psline[arrows=->](11.4,4)(12.4,4)%
\pspolygon[fillcolor=ocker,fillstyle=solid](12.4,3.4)(12.4,4.6)(13.4,4)%
\psline[arrows=->](13.4,4)(14.4,4)%
\rput(2.3,5){RF amplifier}%
\rput(4.4,5){Mixer}%
\rput(4.4,0.6){Local oscillator (often locked to a quarz)}%
\rput(6.6,3){Bandpass filter}%
\rput(8.7,5){IF amplifier}%
\rput(10.8,3){Demodulator}%
\rput(12.9,5){Audio amplifier}%
\end{pspicture}%
%
\caption{Schematic drawing of a superheterodyne receiver}%
\label{superhet}%
\end{figure}
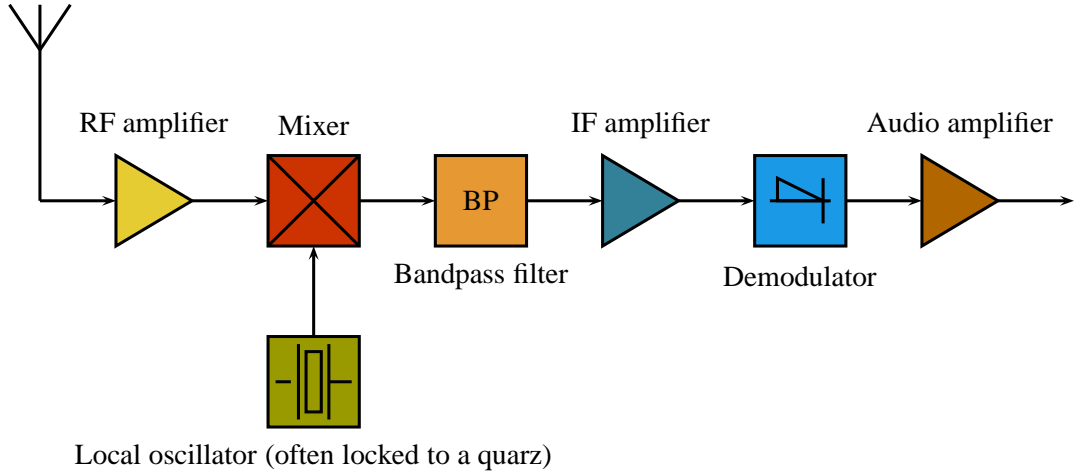

The word superheterodyne is composed of three parts: super (Latin: over), $\epsilon \tau \epsilon \rho \omega$ (hetero, Greek: different) and $\delta \upsilon \nu \alpha \mu \iota \sigma$ (dynamis, Greek: force) and can be translated as two forces superimposed\footnote{The direct translation (roughly) would be: Another force becomes superimposed.}. Different abbrevations exist for the superheterodyne concept. In the US it is often referred to by the simple word `heterodyne' and in Germany one can find the terms `super' or `superhet'. 
The `weak' incident signal is subjected to nonlinear superposition (i.e., mixing or multiplication) with a `strong' sine wave from a local oscillator. At the mixer output we then get the sum and difference frequencies of the signal and local oscillator. The LO signal can be tuned such that the output signal is always at the same frequency or in a very narrow frequency band. Therefore a fixed frequency bandpass with excellent transfer characteristics can be used which is cheaper and easier than a variable bandpass with the same performance. A well-known application of this principle is any simple radio receiver (Fig.\,\ref{superhet}).
%
%
\section{Oscilloscope}
An oscilloscope is typically used for acquisition, display, and measurement of signals in time domain. 
The bandwidth of real-time oscilloscopes is limited in most cases to 10\,GHz. 
For higher bandwidth on repetitive signals the sampling technique has been in use since about 1960. One of the many interesting features of modern oscilloscopes is that they can change the sampling rate through the sweep in a programmed manner. This can be very helpful for detailed analysis in certain time windows. Typical sampling rates are between a factor 2.5 and 4 of the maximum frequency (according to the Nyquist theorem a real-time minimum sampling rate of twice the maximum frequency $f_{\text{max}}$ is required).

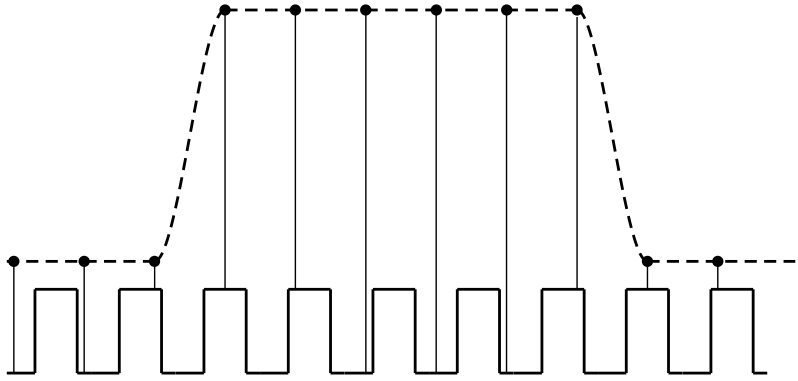
\begin{figure}[htbp]%
\centering
\resizebox{0.7\textwidth}{!}{\begin{pspicture}(0,0)(12,6)%
\psset{linewidth=0.04}%
\psplot[algebraic=true,linestyle=dashed]{2.5}{3.5}{-7.2*x*x*x+64.8*x*x-189*x+182}
\rput(6,7.6){\psplot[algebraic=true,linestyle=dashed]{2.5}{3.5}{7.2*x*x*x-64.8*x*x+189*x-182}}%
\psline[linestyle=dashed](0.4,2)(2.5,2)%
\psline[linestyle=dashed](3.5,5.6)(8.5,5.6)%
\psline[linestyle=dashed](9.5,2)(11.6,2)%
\multirput(0,0)(1.2,0){9}{%
\psline(0.4,0.4)(0.8,0.4)%
\psline(0.8,0.4)(0.8,1.6)%
\psline(0.8,1.6)(1.4,1.6)%
\psline(1.4,1.6)(1.4,0.4)%
\psline(1.4,0.4)(1.6,0.4)%
}%
\multirput(0,0)(1,0){3}{\pscircle[fillstyle=solid,fillcolor=black](0.5,2){0.08}}%
\multirput(0,0)(1,0){6}{\pscircle[fillstyle=solid,fillcolor=black](3.5,5.6){0.08}}%
\multirput(0,0)(1,0){2}{\pscircle[fillstyle=solid,fillcolor=black](9.5,2){0.08}}%
\psset{linewidth=0.02}%
\multirput(0,0)(1,0){2}{\psline(0.5,2)(0.5,0.4)}%
\psline(2.5,2)(2.5,1.6)%
\multirput(0,0)(1,0){2}{\psline(3.5,5.6)(3.5,1.6)}%
\multirput(0,0)(1,0){3}{\psline(5.5,5.6)(5.5,0.4)}%
\psline(8.5,5.5)(8.5,1.6)
\multirput(0,0)(1,0){2}{\psline(9.5,2)(9.5,1.6)}%
\end{pspicture}}%
\caption{Illustration of sequential sampling}%
\label{sequ}%
\end{figure}

Sequential sampling (Fig.\,\ref{sequ}) requires a pre-trigger (required to open the sampling gate) and permits  a non-real-time bandwidth of more than 100\,GHz with modern scopes.

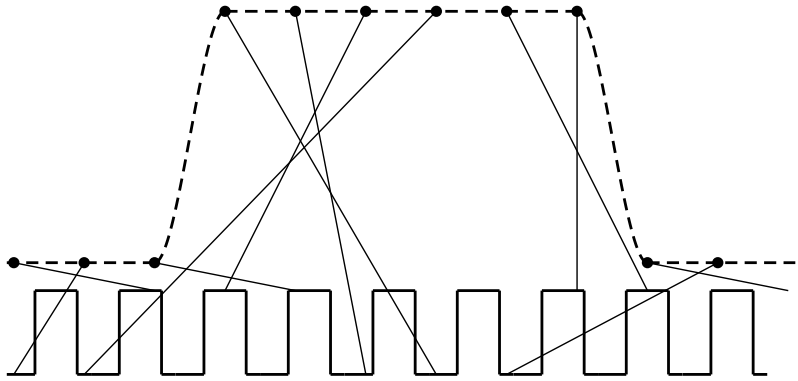
\begin{figure}[htbp]%
\centering
\resizebox{0.7\textwidth}{!}{\begin{pspicture}(0,0)(12,6)%
\psset{linewidth=0.04}%
\psplot[algebraic=true,linestyle=dashed]{2.5}{3.5}{-7.2*x*x*x+64.8*x*x-189*x+182}
\rput(6,7.6){\psplot[algebraic=true,linestyle=dashed]{2.5}{3.5}{7.2*x*x*x-64.8*x*x+189*x-182}}%
\psline[linestyle=dashed](0.4,2)(2.5,2)%
\psline[linestyle=dashed](3.5,5.6)(8.5,5.6)%
\psline[linestyle=dashed](9.5,2)(11.6,2)%
\multirput(0,0)(1.2,0){9}{%
\psline(0.4,0.4)(0.8,0.4)%
\psline(0.8,0.4)(0.8,1.6)%
\psline(0.8,1.6)(1.4,1.6)%
\psline(1.4,1.6)(1.4,0.4)%
\psline(1.4,0.4)(1.6,0.4)%
}%
\multirput(0,0)(1,0){3}{\pscircle[fillstyle=solid,fillcolor=black](0.5,2){0.08}}%
\multirput(0,0)(1,0){6}{\pscircle[fillstyle=solid,fillcolor=black](3.5,5.6){0.08}}%
\multirput(0,0)(1,0){2}{\pscircle[fillstyle=solid,fillcolor=black](9.5,2){0.08}}%
\psset{linewidth=0.02}%
\psline(0.5,0.4)(1.5,2)
\psline(1.5,0.4)(6.5,5.6)
\psline(2.5,1.6)(0.5,2)
\psline(3.5,1.6)(5.5,5.6)
\psline(4.5,1.6)(2.5,2)
\psline(5.5,0.4)(4.5,5.6)
\psline(6.5,0.4)(3.5,5.6)
\psline(7.5,0.4)(10.5,2)
\psline(8.5,1.6)(8.5,5.6)
\psline(9.5,1.6)(7.5,5.6)
\psline(11.5,1.6)(9.5,2)
\end{pspicture}}%
\caption{Illustration of random sampling}%
\label{random}%
\end{figure}

Random sampling (rarely used these days, Fig.\,\ref{random}) was developed about 40 years ago (around 1970) for the case where no pre-trigger was available and relying on a strictly periodic signal to predict a pre-trigger from the measured periodicity.

Sampling is discussed in more detail in the following. Consider a bandwidth-limited time function s($t$) and its Fourier transform S($f$). The signal s($t$) is sampled (multiplied) by a series of equidistant $\delta$-pulses p($t$) \cite{luke}:
\begin{equation}
\text{p}(t) = \sum^{+\infty}_{n=-\infty}{\delta (t - nT_{\text{s}}) = III(t/T_{\text{s}})}
\label{equidistdelta}
\end{equation}
where the symbol $III$ is derived from the Russian letter III and is pronounced `sha'. It represents a series of $\delta$-pulses.

\begin{figure}[htbp]%
\centering
\input{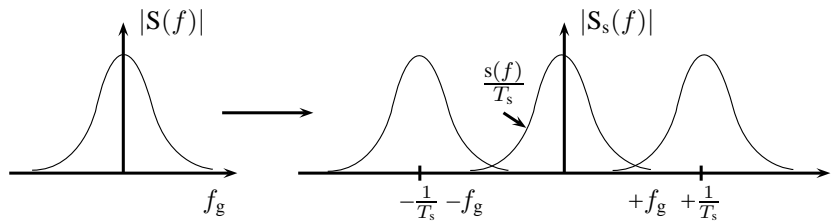}%
\caption{Periodically repeated component of the Fourier Transform of s$_{\text{s}}(t)$ \cite{luke}}%
\label{periodic}%
\end{figure}

The sampled time functions s$_{\text{s}}(t)$ is
\begin{eqnarray}
\nonumber \text{s}_{\text{s}}(t) = \text{s}(t)\text{p}(t) = \text{s}(t)III(t/T_{\text{s}}) \\
\text{S}_{\text{s}}(f) = \text{S}(f)*\frac{1}{T_{\text{s}}}III(T_{\text{s}}f) \\
\text{S}_{\text{s}}(f) = \frac{1}{T_{\text{s}}}\sum^{+\infty}_{n=-\infty}{\text{S}(f - mF)} \text{ with } F = \frac{1}{T_{\text{s}}}\hspace{0.2cm}.
\label{sampledtimefunc}
\end{eqnarray}
Note that the spectrum is repeated periodically by the sampling process. For proper reconstruction, one ensures that overlapping as in Fig.\,\ref{periodic} does not occur.

\begin{figure}[htbp]%
\centering
\input{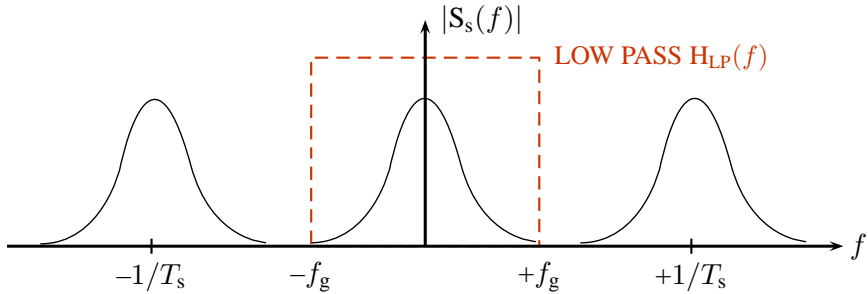}%
\caption{Reconstruction of S$(f)$ via ideal lowpass from S$_{\text{s}}(f)$ (slightly oversampled)}%
\label{oversampled}%
\end{figure}

If the spectra overlap as in Fig.\,\ref{periodic} we have undersampling, the sampling rate is too low. If big gaps occur between the spectra (Fig.\,\ref{oversampled}) we have oversampling, the sampling rate is too high. 
But this scheme applies in most cases. In the limit we arrive at a Nyquist rate of $1/T_{\text{s}} = 2f_{\text{g}} = F$.

The rules mentioned above are of great importance for all `digital' oscilloscopes. The performance (conversion time, resolution) of the input ADC (analog--digital converter) is the key element for single-shot rise time. With several ADCs in time-multiplex one obtains these days 8-bit vertical resolution at 20\,GSa/s = 10\,GHz bandwidth.

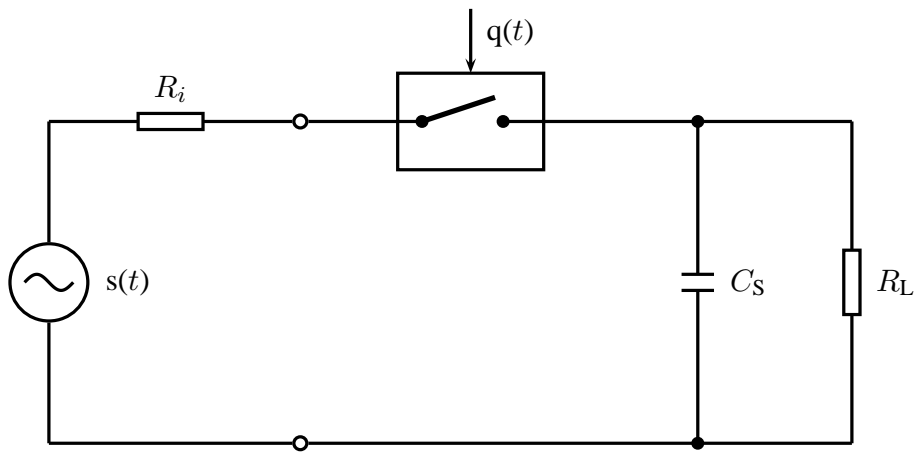
\begin{figure}[htbp]%
\centering
\resizebox{0.8\textwidth}{!}{\begin{pspicture}(0,0)(12,7.4)%
\psset{linewidth=0.04}%
\psline(1,1)(1,2.5)
\pscircle(1,3){0.5}
\rput(0.7,3){\psplot[algebraic=true]{0}{0.6}{0.1*sin(10.4719755*x)}}
\psline(1,3.5)(1,5)
\psline(1,1)(4,1)
\psline(1,5)(2.1,5)
\pspolygon(2.1,4.9)(2.9,4.9)(2.9,5.1)(2.1,5.1)
\psline(2.9,5)(4,5)
\pscircle(4.1,5){0.1}
\pscircle(4.1,1){0.1}
\psline(4.2,1)(10.9,1)
\psline(4.2,5)(5.6,5)
\pspolygon(5.3,4.4)(7.1,4.4)(7.1,5.6)(5.3,5.6)
\psline[linewidth=0.065](5.6,5)(6.5,5.3)
\pscircle[fillstyle=solid,fillcolor=black](5.6,5){0.08}
\pscircle[fillstyle=solid,fillcolor=black](6.6,5){0.08}
\psline(6.6,5)(10.9,5)
\psline[arrows=->](6.2,6.4)(6.2,5.6)
\psline(9,1)(9,2.9)%
\psline(8.8,2.9)(9.2,2.9)%
\psline(8.8,3.1)(9.2,3.1)%
\psline(9,3.1)(9,5)%
\pscircle[fillstyle=solid,fillcolor=black](9,1){0.08}%
\pscircle[fillstyle=solid,fillcolor=black](9,5){0.08}%
\psline(10.9,1)(10.9,2.6)%
\pspolygon(10.8,2.6)(11,2.6)(11,3.4)(10.8,3.4)%
\psline(10.9,3.4)(10.9,5)%
\rput[l](1.7,3){s($t$)}%
\rput(2.5,5.4){$R_{i}$}%
\rput[l](6.4,6.1){q($t$)}%
\rput[l](9.4,3){$C_{\text{S}}$}%
\rput[l](11.2,3){$R_{\text{L}}$}%
\end{pspicture}%
}%
\caption{Sampling gate as harmonic mixer; C$_{\text{s}}$ = sampling capacitor \cite{Schiek2}}%
\label{harmonmix}%
\end{figure}

Another way to look at the sampling theorem (Nyquist) is to consider the sampling gate as a harmonic mixer (Fig.\,\ref{harmonmix}).
%
%

This is basically a nonlinear element (e.g., a diode) that gives product terms of two signals superimposed on its nonlinear characteristics.

The switch in Fig.\,\ref{harmonmix} may be considered as a periodically varying resistor R($t$) actuated by q($t$). If q($t$) is not exactly a $\delta$-function then the higher harmonics decrease with $f$ and the spectral density becomes smaller at high frequencies.

\begin{figure}[htbp]%
\centering
\begin{pspicture}(0,0)(12,2.8)%
\psset{linewidth=0.04}%
\multirput(0.3,1.4)(2.4,0){5}{
\psplot[algebraic=true]{0.45}{1.5}{sin(6*x + 3.14192)}
\psline(0.3,-0.4)(1.7,-0.4)%
}%
\rput(-3.7,1.4){\psplot[algebraic=true,linestyle=dashed]{4.5}{15}{0.6*sin(0.6*x + 3.14192)}}
\pscircle[fillstyle=solid,fillcolor=black](0.79,1.14){0.08}
\pscircle[fillstyle=solid,fillcolor=black](3.32,1.93){0.08}
\pscircle[fillstyle=solid,fillcolor=black](6.1,1.63){0.08}
\pscircle[fillstyle=solid,fillcolor=black](8.63,0.87){0.08}
\pscircle[fillstyle=solid,fillcolor=black](11,1.05){0.08}
\psset{linewidth=0.02}%
\psline(0.8,1)(0.8,0.8)%
\psline(3.4,1)(3.4,0.8)%
\psline(6,1)(6,0.8)%
\psline(8.8,1)(8.8,0.8)%
\psline(11.4,1)(11.4,0.8)%
\rput(3.4,0.2){\tiny $\Delta t$}%
\rput(6,0.2){\tiny 2 $\Delta t$}%
\rput(8.8,0.2){\tiny 3 $\Delta t$}%
\rput(11.4,0.2){\tiny 4 $\Delta t$}%
\end{pspicture}
\caption{Signal reconstruction with sampling shift by $\Delta t$ per pulse \cite{Lipinski}}%
\label{reconstr}%
\end{figure}
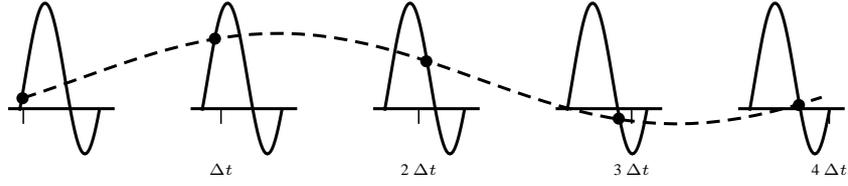

For periodic signals one may apply a special sampling scheme. With each signal event the sampling time is moved by a small fraction $\Delta t$ along the signal to be measured (Fig.\,\ref{reconstr}). 
The highest possible signal frequency for this sequential sampling is linked to the width of the sampling pulse. This sampling or gating pulse should be as short as possible otherwise signal averaging during the `gate-open' period would take place.

\begin{figure}[htbp]%
\centering
\resizebox{0.9\textwidth}{!}{\begin{pspicture}(0,0)(15.5,6)%
\psset{linewidth=0.04}%
%
%
\psline(1,1)(1,2.5)
\pscircle(1,3){0.5}
\rput(0.7,3){\psplot[algebraic=true]{0}{0.6}{0.1*sin(10.4719755*x)}}
\psline(1,3.5)(1,5)
\psline(1,1)(7.7,1)
\psline(1,5)(2.1,5)
\psline(2.9,5)(7.7,5)
\pspolygon[fillstyle=solid](2.1,4.9)(2.9,4.9)(2.9,5.1)(2.1,5.1)
\psline(1,5)(7.7,5)
\pspolygon(4,4.4)(5.8,4.4)(5.8,5.6)(4,5.6)
\pspolygon[fillstyle=solid,fillcolor=black](4.7,4.8)(4.7,5.2)(5.2,5)
\psline(5.2,4.8)(5.2,5.2)
\psline[arrows=->](4.9,3.6)(4.9,4.4)
\pspolygon[fillstyle=solid](2.1,4.9)(2.9,4.9)(2.9,5.1)(2.1,5.1)
\psline(6.8,1)(6.8,2.9)%
\psline(6.6,2.9)(7,2.9)%
\psline(6.6,3.1)(7,3.1)%
\psline(6.8,3.1)(6.8,5)%
\pscircle[fillstyle=solid,fillcolor=black](6.8,1){0.08}%
\pscircle[fillstyle=solid,fillcolor=black](6.8,5){0.08}%
\pscircle[fillstyle=solid](7.7,1){0.1}
\pscircle[fillstyle=solid](7.7,5){0.1}
\psline[arrows=->](7.7,4.7)(7.7,1.3)%
\rput[l](1.7,3){s($t$)}%
\rput(2.5,5.4){$R_{i}$}%
\rput[l](5.1,4){q($t$)}%
\rput[l](7.2,3){$C_{\text{S}}$}%
\rput[l](7.9,3){u$_{\text{C}}(t)$}%
\pspolygon[fillstyle=solid](9,2.9)(9.8,2.9)(9.8,2.8)(10,3)(9.8,3.2)(9.8,3.1)(9,3.1)%
%
%
\rput(10,0){%
\psline(1,1)(1,2.5)
\pscircle(1,3){0.5}
\rput(0.7,3){\psplot[algebraic=true]{0}{0.6}{0.1*sin(10.4719755*x)}}
\psline(1,5)(2.1,5)
\pspolygon[fillstyle=solid](2.1,4.9)(2.9,4.9)(2.9,5.1)(2.1,5.1)
\rput[l](1.7,3){s($t$)}%
\rput(2.5,5.4){$R_{\text{d}}(t)$}%
}%
\psline(11,3.5)(11,3.85)%
\pspolygon[fillstyle=solid](10.9,3.85)(11.1,3.85)(11.1,4.65)(10.9,4.65)
\psline(11,4.65)(11,5)%
\psline(11,1)(15,1)%
\psline(13,5)(15,5)%
\rput(7.4,0){%
\psline(6.8,1)(6.8,2.9)%
\psline(6.6,2.9)(7,2.9)%
\psline(6.6,3.1)(7,3.1)%
\psline(6.8,3.1)(6.8,5)%
\pscircle[fillstyle=solid,fillcolor=black](6.8,1){0.08}%
\pscircle[fillstyle=solid,fillcolor=black](6.8,5){0.08}%
\pscircle[fillstyle=solid](7.7,1){0.1}
\pscircle[fillstyle=solid](7.7,5){0.1}
\rput[l](7.2,3){$C_{\text{S}}$}%
}%
\rput(11.3,4.25){$R_{i}$}%
\end{pspicture}%
}%
\caption{Equivalent circuit for the sample-and-hold element \cite{Schiek2}}%
\label{sampleandhold}%
\end{figure}
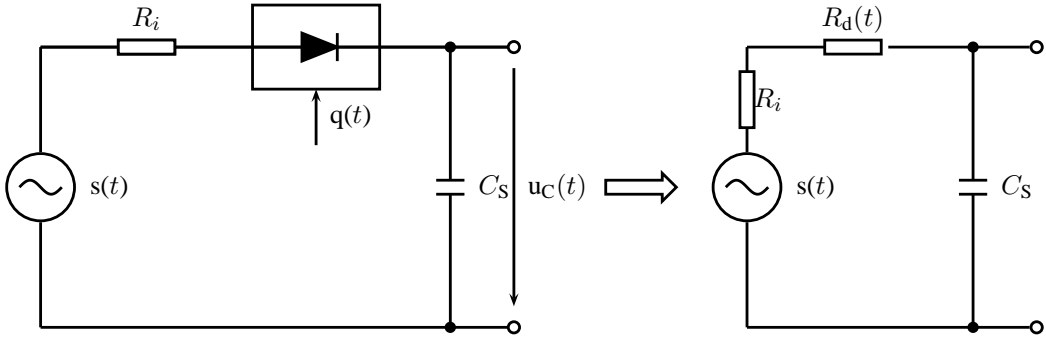

The sampling pulse is often generated by step-recovery diodes (snap-off diodes) which change their conductivity very rapidly between the conducting and non-conducting state. The actual switch (Schottky diode) becomes conductive during the gate pulse and charges a capacitor (sample and hold circuit) but not to the full signal voltage. Assuming a time constant $R_{i}C_{\text{s}}$ much bigger than the `open' time of the sampling gate, we obtain approximately (Fig.\,\ref{sampleandhold})
\begin{equation}
i_{\text{c}}(t) = \frac{\text{s}(t)}{R_{i} + R_{\text{d}}(t)}\hspace{0.2cm}.
\label{samplehold}
\end{equation}
%

\begin{figure}[htbp]%
\centering
\input{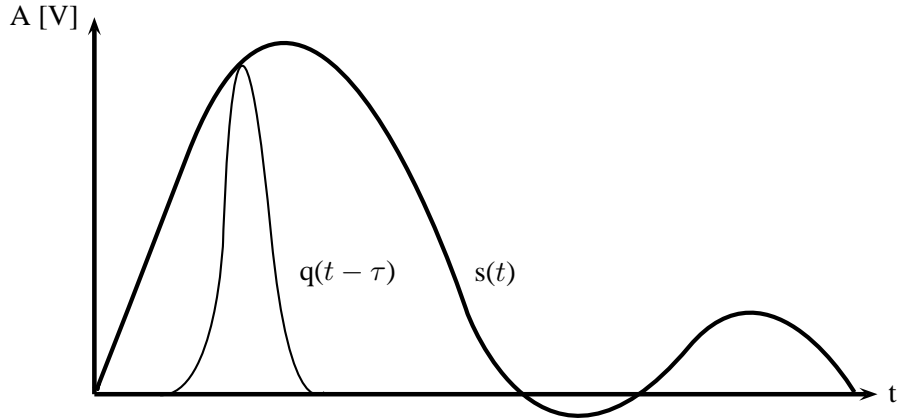}%
\caption{Sampling with finite-width sampling pulse}%
\label{sampling}%
\end{figure}

After the sampling process we have \cite{Schiek2}
\begin{equation}
u_{\text{c}}(t) = \frac{1}{C_{\text{s}}}\int^{+\infty}_{-\infty} i_{\text{c}}(t)\text{d}t = \int^{+\infty}_{-\infty} \text{s}(t)\frac{1}{C_{\text{s}}(R_{i} + R_{\text{d}}(t))}\hspace{0.2cm},
\label{aftersampling}
\end{equation}
with the control signal for the Schottky diode being
\begin{equation}
\text{q}(t) = \frac{1}{C_{\text{s}}(R_{i} + R_{\text{d}}(t))}\hspace{0.2cm}.
\label{controlsignal}
\end{equation}
The control or switching signal is moved by $\tau$ or n$\Delta t$ (Fig.\,\ref{timing}) with respect to the signal to be sampled s($t$):
\begin{equation}
u_{\text{c}}(\tau) = \int^{+\infty}_{-\infty}{\text{s}(t)\text{q}(t - \tau)\text{d}t}\hspace{0.2cm}.
\label{switch}
\end{equation}
Note that the time constant $R_{i}C_{\text{s}}$ is much bigger than the length of q$(t)$. $C_{\text{s}}$ is only charged to a fraction of s$(t)$ (Fig.\,\ref{sampling}).
The sampling efficiency $\eta$ is defined as
\begin{equation}
\eta = \frac{u_{\text{c}}(\tau)}{\text{s}(\tau)}\hspace{0.2cm}.
\label{efficiency}
\end{equation}
In order to circumvent the problem of poor sampling efficiency a feedback loop technique (integrator) can be used. This integrator amplifies the voltage step on the sampling capacitor, after the sampling gate is closed, exactly by a factor $1/\eta$. If the sampling gate has not moved with respect to the trigger, the sampling capacitor is already charged to the correct voltage u$_{c}(\tau)$ and there is no change. Otherwise the change in u$_{c}$ just amounts to the change in signal voltage.

The sampling gate is interesting from a technological point of view. As aperture times (Fig.\,\ref{sampling}) may be of the order of 10\,ps, MIC (Microwave Integrated Circuit) technology has been used for many years. Today, the latest generation of sampling heads (50\,GHz) is even one step further with MMIC (Monolithic Microwave Integrated Circuits) technology.

\begin{figure}[htbp]%
\centering
\includegraphics[width=\mywidth]{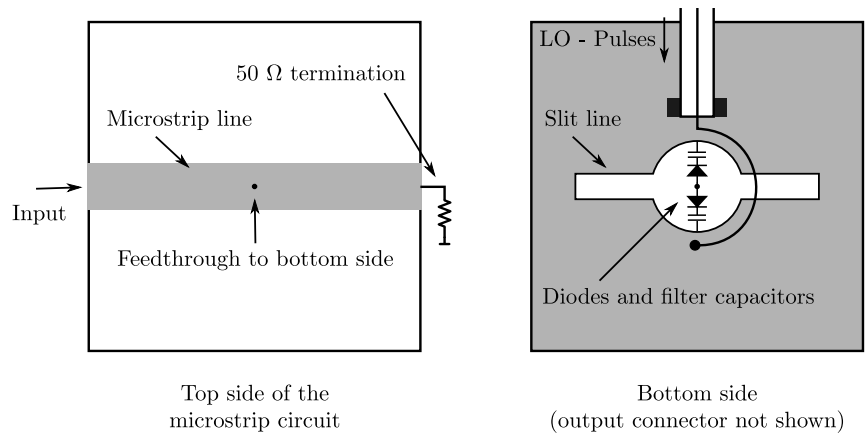}%
\caption{Sampling circuits \cite{Lipinski}}%
\label{circuits}%
\end{figure}

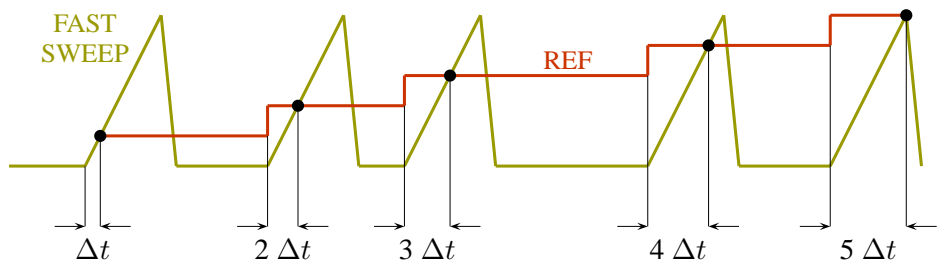
\begin{figure}[htbp]%
\centering
\begin{pspicture}(0,0)(12,3.5)%
\newrgbcolor{myred}{0.8 0.2 0}%
\newrgbcolor{olive}{0.6 0.6 0}%
\psset{linewidth=0.04,linecolor=olive}%
\rput(0,0.5){%
\psline(1,0.8)(2,2.8)
\psline(2,2.8)(2.2,0.8)
\rput(2.4,0){%
\psline(1,0.8)(2,2.8)
\psline(2,2.8)(2.2,0.8)
}%
\rput(4.2,0){%
\psline(1,0.8)(2,2.8)
\psline(2,2.8)(2.2,0.8)
}%
\rput(7.4,0){%
\psline(1,0.8)(2,2.8)
\psline(2,2.8)(2.2,0.8)
}%
\rput(9.8,0){%
\psline(1,0.8)(2,2.8)
\psline(2,2.8)(2.2,0.8)
}%
\psline(0,0.8)(1,0.8)%
\psline(2.2,0.8)(3.4,0.8)%
\psline(4.6,0.8)(5.2,0.8)%
\psline(6.4,0.8)(8.4,0.8)%
\psline(9.6,0.8)(10.8,0.8)%
\psset{linecolor=myred}%
\psline(1.2,1.2)(3.4,1.2)%
\psline(3.4,1.6)(5.2,1.6)%
\psline(5.2,2)(8.4,2)%
\psline(8.4,2.4)(10.8,2.4)%
\psline(10.8,2.8)(11.8,2.8)%
\psline(3.4,1.2)(3.4,1.6)%
\psline(5.2,1.6)(5.2,2)%
\psline(8.4,2)(8.4,2.4)%
\psline(10.8,2.4)(10.8,2.8)%
\psset{linecolor=black}%
\pscircle[fillstyle=solid,fillcolor=black](1.2,1.2){0.08}%
\pscircle[fillstyle=solid,fillcolor=black](3.8,1.6){0.08}%
\pscircle[fillstyle=solid,fillcolor=black](5.8,2){0.08}%
\pscircle[fillstyle=solid,fillcolor=black](9.2,2.4){0.08}%
\pscircle[fillstyle=solid,fillcolor=black](11.8,2.8){0.08}%
\psset{linewidth=0.01}%
\psline(1,0)(1,0.8)%
\psline(1.2,0)(1.2,1.2)%
\psline(3.4,0)(3.4,1.2)%
\psline(3.8,0)(3.8,1.6)%
\psline(5.2,0)(5.2,1.6)%
\psline(5.8,0)(5.8,2)%
\psline(8.4,0)(8.4,2)%
\psline(9.2,0)(9.2,2.4)%
\psline(10.8,0)(10.8,2.4)%
\psline(11.8,0)(11.8,2.8)%
\psset{arrowsize=0.1}%
\psline[arrows=->](0.6,0)(1,0)
\rput(2.4,0){%
\psline[arrows=->](0.6,0)(1,0)%
}%
\rput(4.2,0){%
\psline[arrows=->](0.6,0)(1,0)%
}%
\rput(7.4,0){%
\psline[arrows=->](0.6,0)(1,0)%
}%
\rput(9.8,0){%
\psline[arrows=->](0.6,0)(1,0)%
}%
\psline[arrows=->](1.6,0)(1.2,0)
\rput(2.6,0){%
\psline[arrows=->](1.6,0)(1.2,0)%
}%
\rput(4.6,0){%
\psline[arrows=->](1.6,0)(1.2,0)%
}%
\rput(8,0){%
\psline[arrows=->](1.6,0)(1.2,0)%
}%
\rput(10.6,0){%
\psline[arrows=->](1.6,0)(1.2,0)%
}%
}%
\rput(1.1,0.2){$\Delta t$}%
\rput(3.6,0.2){2 $\Delta t$}%
\rput(5.5,0.2){3 $\Delta t$}%
\rput(8.8,0.2){4 $\Delta t$}%
\rput(11.3,0.2){5 $\Delta t$}%
\rput(1,3.2){\small \olive FAST}%
\rput(1,2.8){\small \olive SWEEP}%
\rput(7.35,2.7){\small \myred REF}%
\end{pspicture}%
\caption{Timing of sampling pulses \cite{Lipinski}}%
\label{timing}%
\end{figure}

In MIC technology the sampling pulse is applied to a slotline in the ground-plane metallization of a microstrip substrate (Fig.\,\ref{circuits}). 
This slot line has a length of some 10\,mm and is shorted at both ends. With a voltage across the slotline the fast Schottky diodes open and connect the microstrip line via a through hole to the sampling capacitor $C_{\text{s}}$. Owing to the particular topology of the circuit the signal line (microstrip) is decoupled from the sampling pulse line over a wide frequency range (Fig.\,\ref{circuits}).

To move the sampling pulse by $\Delta t$ for each event requires a pre-trigger (several 10 ns ahead), to start a fast-ramp generator. The intersection (comparator) of the ramp generator output with a staircase-like reference voltage defines the sampling time and $\Delta t$ (Fig.\,\ref{timing}).

The delay required for the pre-trigger has been a significant problem for many applications, since it may be as large as 70\,ns on certain (older) instruments. A 70\,ns delay-line leads to considerable signal distortions especially for the high-frequency components.

To avoid the delay for the pre-trigger a technique named `random sampling' was developed about 45 years ago. It requires a strictly periodic signal rather than just the repetitive one for sequential sampling (Fig.\,\ref{timing}). By measuring the (constant) repetition frequency of this strictly periodic signal, a prediction of the next pulse arrival time can be given in order to generate a trigger. Today there is little interest in random sampling, as pre-trigger delays are drastically reduced (12 ns). There are also problems with jitter, and random sampling needs repetition rates of serveral kHz \cite{Lipinski}.

Features of modern sampling scopes are summarized in Table \ref{scopefeat}.
\begin{table}
\centering
\caption{Features of modern sampling scopes}
\begin{tabular}{lc}
\hline
\hline
Rise time: 7 ps & $\approx$ 50 GHz \\
Jitter & $\approx$ 1.5 ps \\
Static operation possible, no minimum repetition rate required. & \\
Optical sampling (mode-locked laser), 1 ps rise time & $\approx$ 350 GHz \\
\hline
\hline
\end{tabular}
\label{scopefeat}
\end{table}
\section{Spectrum analyser}
Radio-frequency spectrum analysers can be found in virtually every control-room of a modern particle accelerator. They are used for many aspects of beam diagnostics including Schottky signal acquisition and RF observation. A spectrum analyser is in principle very similar to a common superheterodyne broadcast receiver, except for the requirements of choice of functions and change of parameters. It sweeps automatically through a specified frequency range which corresponds to an automatic turning of the nob on a radio. The signal is then displayed in the amplitude/frequency plane.
Thirty years ago, instruments were set manually and had some sort of analog or CRT (cathode ray tube) display. Nowadays, with the availability of cheap and powerful digital electronics for control and data processing, nearly all instruments can be remotely controlled. The microprocessor permits fast and reliable setting of the instrument and reading of the measured values. Extensive data treatment for error correction, complex calibration routines, and self tests are a great improvement. However, the user of such a sophisticated system may not always be aware what is really going on in the analog section before all data are digitized. The basis of these analog sections are discussed now.

In general there are two types of spectrum analyser:
\begin{itemize}
	\item Scalar spectrum analysers (SA) and
	\item Vector spectrum analysers (VSA).
\end{itemize}
The SA provides only information of the amplitude of an ingoing signal, while the VSA provides the phase as well.
\subsection{Scalar spectrum analysers}
A common oscilloscope displays a signal in the amplitude-time plane (time domain). The SA follows another approach and displays it in frequency domain. 

\begin{figure}[htbp]%
\centering
\begin{pspicture}(0,0)(12.5,5)%
\newrgbcolor{orange}{0.9 0.6 0.2}%
\psset{linewidth=0.04}%
\rput(1,1){\psplot[algebraic=true,linecolor=orange]{0}{4}{0.06*sin(5*x)}}
\rput(1,4){\psplot[algebraic=true,linecolor=orange]{0}{4}{0.06*sin(5*x)}}
\psline[linewidth=0.08,linecolor=orange](9.4,0.6)(9.5,4)
\psline[linewidth=0.08,linecolor=orange](9.6,0.6)(9.5,4)
\psline[linewidth=0.08,linecolor=orange](7.9,0.6)(8,2)
\psline[linewidth=0.08,linecolor=orange](8.1,0.6)(8,2)
\psline[linewidth=0.08,linecolor=orange](10.9,0.6)(11,2)
\psline[linewidth=0.08,linecolor=orange](11.1,0.6)(11,2)
\psline[arrows=->](1,0.6)(1,4.4)
\psline[arrows=->](1,2.5)(5.5,2.5)
\psline[arrows=->](7,0.6)(7,4.4)
\psline[arrows=->](7,0.6)(12,0.6)
\psline(6.9,2)(7.1,2)%
\psline(6.9,4)(7.1,4)%
\rput[r](0.8,4.4){A [V]}%
\rput[r](6.8,4.4){A [dB]}%
\rput[l](5.7,2.5){t}%
\rput[l](12.2,0.6){t}%
\rput(3.25,0.2){\small 2\%\,AMPLITUDE MODULATION}%
\rput(9.5,0.2){\small 2\%\,AM IN FREQUENCY DOMAIN}%
\rput[r](6.8,2){\small -- 40}%
\rput[r](6.8,4){\small 0}%
\end{pspicture}%
\caption{Example of amplitude modulation in time and frequency domain}%
\label{am}%
\end{figure}
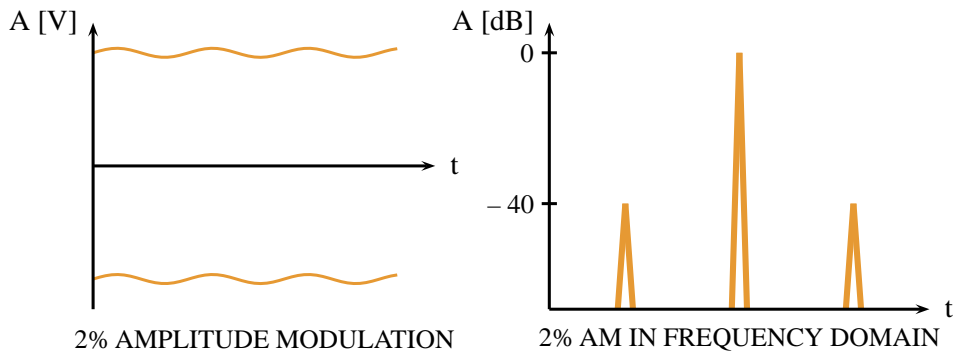

One of the major advantages of the frequency-domain display is the sensitivity to periodic perturbations. For example, 5\% distortion is already difficult to see in the time domain but in the frequency domain the sensitivity to such `sidelines' (Fig.\,\ref{am}) is very high ($-$120\,dB below the main line). 
%
A very faint amplitude modulation (AM) of 10$^{-\text{12}}$ (power) on some sinusoidal signals would be completely invisible on the time trace, but can be displayed as two sidelines 120\,dB below the carrier in the frequency domain \cite{Schleifer}.

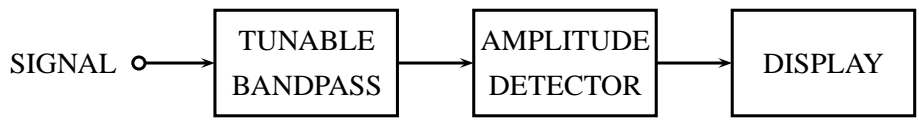
\begin{figure}[htbp]%
\centering
\begin{pspicture}(0,0)(12,2)%
\psset{linewidth=0.04}%
\rput(-0.2,-1){%
\multirput(0,0)(3.4,0){3}{\psline[arrows=->](2,2)(3,2)}%
\multirput(0,0)(3.4,0){3}{\pspolygon(3,1.3)(5.4,1.3)(5.4,2.7)(3,2.7)}%
\pscircle[fillstyle=solid](2,2){0.1}%
\rput[r](1.7,2){SIGNAL}%
\rput(4.2,2.3){TUNABLE}%
\rput(4.2,1.7){BANDPASS}%
\rput(7.6,2.3){AMPLITUDE}%
\rput(7.6,1.7){DETECTOR}%
\rput(11,2){DISPLAY}%
}%
\end{pspicture}%
%
\caption{A tunable bandpass as a simple spectrum analyser (SA)}%
\label{bp}%
\end{figure}

We will now consider only serial processing or swept tuned analysers (Fig.\,\ref{bp}).

The easiest way to design a swept tuned spectrum analyser is by using a tunable bandpass. This may be an LC circuit, or a YIG filter (YIG = Yttrium-Iron-Garnet) beyond 1\,GHz. The LC filter exhibits poor tuning, stability and resolution. YIG filters are used in the microwave range (as preselector) and for YIG oscillators. Their tuning range is about one decade, with Q-values exceeding 1000.

For much better performance the superheterodyne principle can be applied (Fig.\,\ref{superhet}).
\begin{figure}[htbp]%
\centering
\begin{pspicture}(0,0)(13,8)%
\psset{linewidth=0.04}%
\rput(0.5,0){%
\psline(0,6.5)(1,6.5)%
\pscircle[fillstyle=solid](0,6.5){0.1}%
\multirput(0,0)(0,-2.4){3}{\pspolygon(1,5.8)(3.4,5.8)(3.4,7.2)(1,7.2)}%
\rput(3.4,0){\pspolygon(1,5.8)(3.4,5.8)(3.4,7.2)(1,7.2)}%
\rput(8.6,0){\pspolygon(1,5.8)(3.4,5.8)(3.4,7.2)(1,7.2)}%
\rput(8.6,-4.8){\pspolygon(1,5.8)(3.4,5.8)(3.4,7.2)(1,7.2)}%
\rput(5.2,-4.8){\pspolygon(1,5.8)(3.4,5.8)(3.4,7.2)(1,7.2)}%
\pscircle(8.2,6.5){0.5}%
\psline(7.9,6.8)(8.5,6.2)%
\psline(7.9,6.2)(8.5,6.8)%
\pscircle(8.2,4.1){0.5}%
\pspolygon(4.4,1.7)(5.2,1.2)(5.2,2.2)%
\pspolygon(10.8,3.7)(11.3,4.5)(10.3,4.5)%
\psline[arrows=->](3.4,6.5)(4.4,6.5)%
\psline[arrows=->](6.8,6.5)(7.7,6.5)%
\psline[arrows=->](8.7,6.5)(9.6,6.5)%
\psline[arrows=->](10.8,5.8)(10.8,4.5)%
\psline[arrows=->](10.8,3.7)(10.8,2.4)%
\psline[arrows=->](9.6,1.7)(8.6,1.7)%
\psline[arrows=->](6.2,1.7)(5.2,1.7)%
\psline[arrows=->](4.4,1.7)(3.4,1.7)%
\psline[arrows=->](2.2,3.4)(2.2,2.4)%
\psline[arrows=->](3.4,4.1)(7.7,4.1)%
\psline[arrows=->](8.2,4.6)(8.2,6)%
\rput(0,6.9){\small SIGNAL}%
\rput(0,6.1){\small INPUT}%
\rput(2.2,6.8){\small SWITCHABLE}%
\rput(2.2,6.2){\small ATTENUATOR}%
\rput(5.6,6.5){\small LOW PASS}%
\rput(8.2,7.4){\small MIXER}%
\rput(10.8,6.5){\small IF FILTER}%
\rput(2.2,4.5){\small SAW TOOTH}%
\rput(2.2,3.7){\small GENERATOR}%
\rput[r](7.6,5.2){\small TUNABLE}%
\rput[r](7.6,4.6){\small OSCILLATOR}%
\rput(8.2,4.1){\small LO}%
\rput[l](11.6,4.4){\small IF}%
\rput[l](11.6,3.9){\small AMPLIFIER}%
\rput(2.2,1.7){\small DISPLAY}%
\rput(4.8,2.4){\small VIDEO}%
\rput(4.8,1){\small AMPLIFIER}%
\rput(7.4,2){\small VIDEO FILTER}%
\rput(7.4,1.4){\small LOW PASS}%
\rput(10.8,2){\small AMPLITUDE}%
\rput(10.8,1.4){\small DETECTOR}%
}%
\end{pspicture}%
%
\caption{Block diagram of a spectrum analyser}%
\label{sa}%
\end{figure}
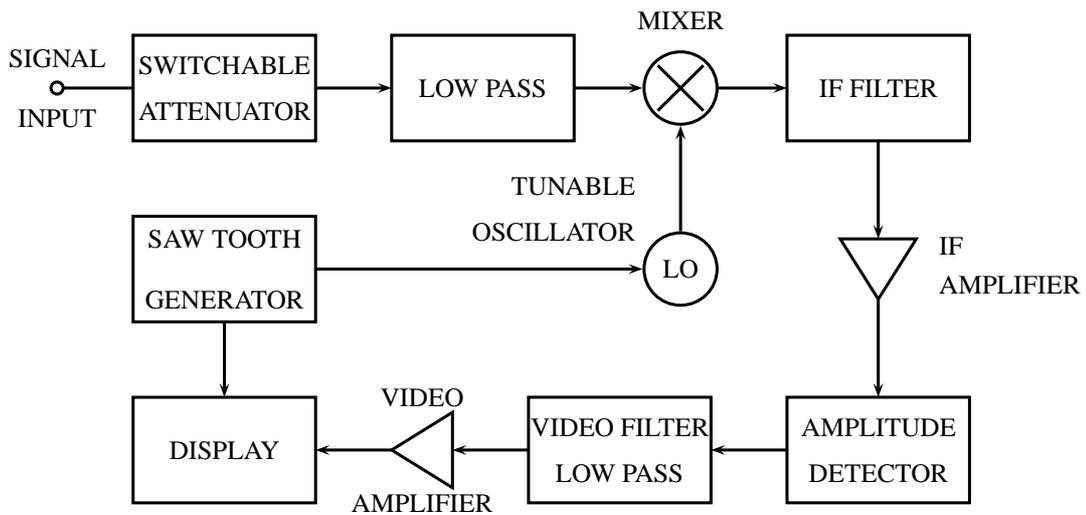

\begin{figure}[htbp]%
\centering
\begin{pspicture}(0,0)(6,6)%
\psset{linewidth=0.04}%
\rput(0.5,0.5){%
\psline[arrows=->](0,0)(0,5.5)%
\psline[arrows=->](0,0)(5,0)%
\psline[linestyle=dashed](0,4)(4.5,5.5)%
\psline(0,0)(5,1.666666)%
\psline[linewidth=0.02](0,5.5)(5,5.5)%
\psline[linewidth=0.02](5,5.5)(5,0)%
}%
\rput(0.2,1.5){1}%
\rput(0.2,2.5){2}%
\rput(0.2,3.5){3}%
\rput(0.2,4.5){4}%
\rput(0.2,5.5){5}%
\rput(0.5,0.2){2}%
\rput[l](5.5,0.2){3 GHZ}%
\rput(3,0.2){f$_{\text{LO}}$}%
\rput(0.8,3.5){f$_{S}$}%
\rput(4.7,5.5){(+)}%
\rput(4.7,2.3){(--)}%
\end{pspicture}%
%
\caption{Frequency chart of the SA of Fig.\,\ref{sa}, intermediate frequency = 2\,GHz}%
\label{fchart}%
\end{figure}
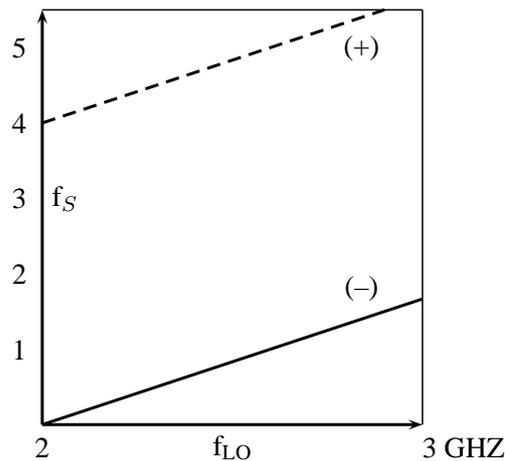

As already mentioned, the nonlinear element (four-diode mixer, double-balanced mixer) delivers mixing products as
\begin{equation}
f_{\text{s}} = f_{\text{LO}} \pm f_{\text{IF}}\hspace{0.2cm}.
\label{eq:23}
\end{equation}
Assuming a signal range from 0 to 1\,GHz for the spectrum analyser depicted in Fig.\,\ref{sa} and $f_{\text{LO}}$ between 2 and 3\,GHz we get the frequency chart shown in Fig.\,\ref{fchart}.

Obviously, for a wide input frequency range without image response we need a sufficiently high intermediate frequency. A similar situation occurs for AM- and FM-broadcast receivers (AM-IF = 455\,kHz, FM-IF = 10.7\,MHz). But for a high intermediate frequency (e.g., 2\,GHz) a stable narrow-band IF filter is difficult to construct which is why most SAs and high quality
receivers use more than one IF. Certain SAs have four different LOs, some fixed, some tunable. For a large tuning range the first, and for a fine tune (e.g., 20\,kHz) the third LO is tuned.

Multiple mixing is necessary when going to a lower intermediate frequency (required when using high-Q quartz filters) for good image response suppression of the mixers.

It can be shown that the frequency of the $n$-th LO must be higher than the (say) 80\,dB bandwidth of the ($n$ $-$ 1)th IF-band filter. A disadvantage of multiple mixing is the possible generation of intermodulation lines if amplitude levels in the conversion chain are not carefully controlled.

The requirements of a modern SA with respect to frequency are
\begin{itemize}
	\item high resolution
	\item high stability (drift, phase noise)
	\item wide tuning range
	\item no ambiguities,
\end{itemize}
and with respect to amplitude response are
\begin{itemize}
	\item large dynamic range (100\,dB)
	\item calibrated, stable amplitude response
	\item low internal distortions.
\end{itemize}
It should be mentioned that the size of the smallest IF-bandpass filter width $\Delta f$ has an important influence on the maximum sweep rate (or step-width and -rate when using a synthesizer)
\begin{equation}
\frac{\text{d}f}{\text{d}t} < (\Delta f)^{2}\hspace{0.2cm}.
\label{eq:24}
\end{equation}
In other words, the signal frequency has to remain at $\Delta T = 1/\Delta f$ within the bandwidth $\Delta f$.

On many instruments the proper relation between $\Delta f$ and the sweep rate is automatically set to the optimum value for the highest possible sweep speed, but it can always be altered manually (setting of the resolution bandwidth).

Certain SAs do not use a sinusoidal LO signal but, rather, periodic short pulses or a comb spectrum (harmonic mixer). This is very closely related to a sampling scope, except that the spacing of the comb lines is different
\begin{equation}
f_{\text{s}} = Nf_{\text{LO}} \pm f_{\text{IF}} \hspace{0.5 cm} n = 1, 2, 3,...
\label{eq:25}
\end{equation}
A single, constant input-frequency line may appear several times on the display. This difficulty (multiple response) was a particular problem with older instruments. Certain modulation and sweep modes permit the identification and rejection of these `ghost' signals. On modern spectrum analysers the problem does not occur, except at frequencies beyond 60\,GHz, when a tracking YIG filter may need to be installed.

Caution is advised when applying, but not necessarily displaying, two or more strong (> 10\,dBm) signals to the input. Intermodulation 3rd-order products may appear (from the first mixer or amplifier) and could lead to misinterpretation of the signals to be analysed.

SAs usually have a rather poor noise figure of 20--40\,dB as they often do not use preamplifiers in front of the first mixer (dynamic range, linearity). But with a good preamplifier the noise figure can be reduced to almost that of the preamplifier. This configuration permits amplifier noise figure measurements to be made with reasonable precision of about 0.5 dB. The input of the amplifier to be tested is connected to a hot and a cold termination and the corresponding two traces on the SA display are evaluated \cite{Schiek, Yip, Evans, Connor, Landstorfer}.

Spectrum analysers can also be used to directly measure the phase noise of an oscillator provided that the LO phase noise in the SA is much lower than that of the device under test \cite{Schiek}. For higher resolution, set-ups with delay lines and additional mixers (SA at low frequencies or FFT) are advised.
\section{Vector spectrum and FFT analyser}
\begin{figure}[htbp]%
\centering
\includegraphics[width=\mywidth]{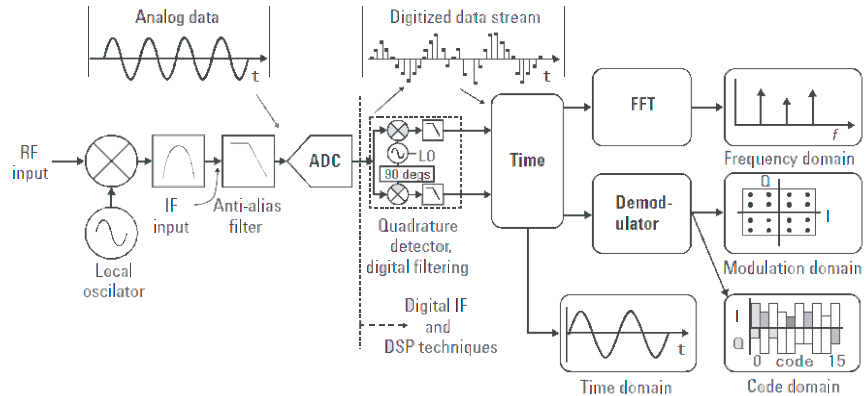}%
\caption{Block diagram of a vector spectrum analyser}%
\label{vsa}%
\end{figure}

The modern vector spectrum analyser (VSA) is essentially a combination of a two-channel digital oscilloscope  and a spectrum analyser FFT display. The incoming signal gets down-mixed, bandpass (BP) filtered and passes an ADC (generalized Nyquist for BP signals; $f_{\text{sample}}$ = 2 BW). A schematic drawing of a modern VSA can be seen in Fig.\,\ref{vsa}.
The digitized time trace then is split into an I (in phase) and Q (quadrature, 90 degree offset) component with respect to the phase of some reference oscillator. Without this reference, the term vector is meaningless for a spectral component.

One of the great advantages is that a VSA can easily separate AM and FM components.

\begin{figure}[htbp]%
\centering
\includegraphics[width=\mywidth]{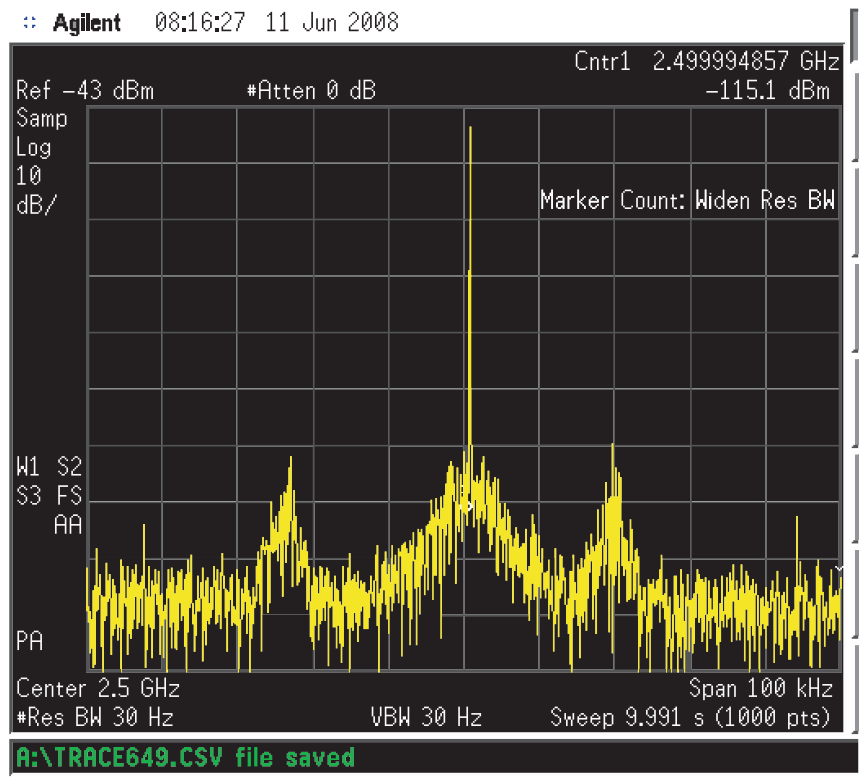}%
\caption{Single-sweep FFT display similar to a very slow scan on a swept spectrum analyser}%
\label{ec1}%
\end{figure}

\begin{figure}[htbp]%
\centering
\includegraphics[width=\mywidth]{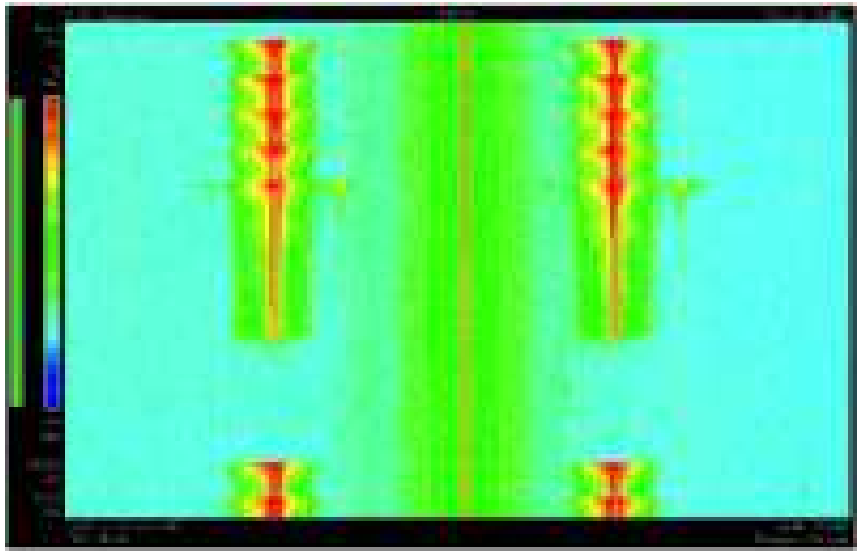}%
\caption{Spectrogram display containing about 200 traces as shown on the left side in colour coding. Time runs from top to bottom}%
\label{ec2}%
\end{figure}

An example of vector spectrum analyser display and performance is given in Fig.\,\ref{ec1} and Fig.\,\ref{ec2}. Both figures were obtained during measurements of the electron cloud in the CERN SPS.
\section{Noise basics}
The concept of `noise' was applied originally to the type of audible sound caused by statistical variations of the air pressure with a wide flat spectrum (white noise). It is now also applied to electrical signals, the noise `floor' determining the lower limit of signal transmission. Typical noise sources are: Brownian movement of charges (thermal noise), variations of the number of charges involved in the conduction (flicker noise), and quantum effects (Schottky noise, shot noise). Thermal noise is only emitted by structures with electromagnetic losses which, by reciprocity, also absorb power. Pure reactances do not emit noise (emissivity = 0).

Different categories of noise can be defined:
\begin{itemize}
	\item white, which has a flat spectrum,
	\item pink, being low-pass filtered, and
	\item blue, being high-pass filtered.
\end{itemize}
In addition to the spectral distribution, the amplitude density distribution is also required in order to characterize a stochastic signal. For signals coming from very many independent sources, the amplitude density has a Gaussian distribution.
The noise power density delivered to a load by a black body is given by Planck's formula:
\begin{equation}
\frac{N_{\text{L}}}{\Delta f} = hf\left(\text{e}^{hf/kT} - 1\right)^{-1}\hspace{0.2cm},
\label{plank}
\end{equation}
where $N_{\text{L}}$ is the noisepower delivered to a load, $h = 6.625 \cdot 10^{-34}$\,Js the Planck constant and $k = 1.38056 \cdot 10^{-23}$\,J/K Boltzmann's constant.

Equation (\ref{plank}) indicates constant noise power density up to about 120\,GHz (at 290\,K) with 1\% error. Beyond, the power density decays and there is no `ultraviolet catastrophe', i.\,e., the total noise power is finite.

The radiated power density of a black body is given as
\begin{equation}
W_{\text{r}}(f,T) = \frac{hf^{3}}{c^{2}\left[\text{e}^{hf/kT} - 1\right]}\hspace{0.2cm}.
\label{radpower}
\end{equation}

For $hf << kT$ the Rayleigh--Jeans approximation of Eq.\,(\ref{plank}) holds:
\begin{equation}
N_{\text{L}} = kT \Delta f\hspace{0.2cm},
\label{rayleighjeans}
\end{equation}
where in this case $N_{\text{L}}$ is the power delivered to a matched load. The no-load noise voltage $u(t)$ of a resistor $R$ is given as
\begin{equation}
\overline{u^{2}(t)} = 4 kT R\Delta f
\label{nlnoise}
\end{equation}
and the short-circuit current $i(t)$ by
\begin{equation}
\overline{i^{2}(t)} = 4 \frac{kT \Delta f}{R} = 4 kT G\Delta f\hspace{0.2cm},
\label{scircurr}
\end{equation}
where $u(t)$ and $i(t)$ are stochastic signals and $G$ is $1/R$. The linear average $\overline{u(t)}, \overline{i(t)}$ vanishes. Of special importance is the quadratic average $\overline{u^{2}(t)}, \overline{i^{2}(t)}$.

\begin{figure}[htbp]%
\centering
\begin{pspicture}(0,0)(12,6)%
\psset{linewidth=0.04}%
\psline(1,1)(1,2.5)%
\pscircle(1,3){0.5}%
\rput(0.7,3){\psplot[algebraic=true]{0}{0.6}{0.1*sin(10.4719755*x)}}%
\psline(1,3.5)(1,5)%
\psline(1,1)(7,1)%
\psline(1,5)(3,5)%
\pspolygon(3,4.9)(3.8,4.9)(3.8,5.1)(3,5.1)%
\psline(3.8,5)(7,5)%
\pscircle(7.1,5){0.1}%
\pscircle(7.1,1){0.1}%
\psline(7.2,1)(9,1)%
\psline(7.2,5)(9,5)%
\psline(9,1)(9,2.6)%
\pspolygon(8.9,2.6)(9.1,2.6)(9.1,3.4)(8.9,3.4)%
\psline(9,3.4)(9,5)%
\rput[l](3.2,5.3){R$_{1}$ = noiseless resistor}%
\rput[l](9.2,3){R$_{2}$ = noiseless load}%
\rput[l](1.6,3){W$_{\text{u}} = 4kT\text{R}_{1}$}%
\end{pspicture}%
\caption{Equivalent circuit for a noisy resistor $R_{1}$ terminated by a noisless load $R_{2}$}%
\label{resistor}%
\end{figure}
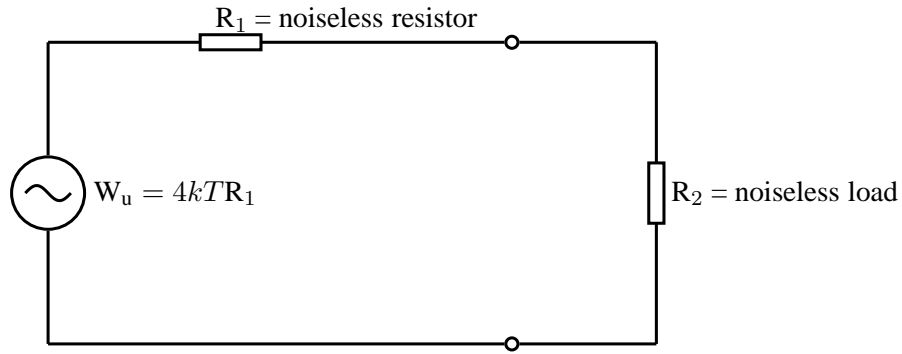

The available power (which is independent of $R$) is given by (Fig.\,\ref{resistor})
\begin{equation}
\frac{\overline{u^{2}(t)}}{4 R} = kT \Delta f\hspace{0.2cm}.
\label{avpower}
\end{equation}

\begin{figure}[htbp]%
\centering
\begin{pspicture}(0,0)(112.5,6)%
\psset{linewidth=0.04}%
\psline[arrows=->](0.2,3)(1,3)%
\psline[linestyle=dashed](1,1)(1,5)
\psline(1,5)(2.2,5)
\pspolygon(2.2,4.9)(3,4.9)(3,5.1)(2.2,5.1)
\psline(3,5)(5,5)
\pscircle(5.5,5){0.5}
\psline(6,5)(12,5)
\psline(1,1)(12,1)
\pscircle[fillstyle=solid](1,1){0.1}
\pscircle[fillstyle=solid](1,5){0.1}
\psline(9,5)(9,4.2)%
\pspolygon(8.9,4.2)(9.1,4.2)(9.1,3.4)(8.9,3.4)%
\psline(9,3.4)(9,2.7)%
\pscircle(9,2.2){0.5}%
\psline(9,1.7)(9,1)%
\pscircle[fillstyle=solid,fillcolor=black](9,5){0.08}%
\pscircle[fillstyle=solid,fillcolor=black](9,1){0.08}%
\rput(3,0){%
\psline(9,5)(9,4.2)%
\pspolygon(8.9,4.2)(9.1,4.2)(9.1,3.4)(8.9,3.4)%
\psline(9,3.4)(9,2.7)%
\pscircle(9,2.2){0.5}%
\psline(9,1.7)(9,1)%
}%
\rput(0.5,3.3){$R_{i},W_{\text{u}}$}%
\rput(2.6,5.3){$R_{1},T_{1}$}%
\rput(5.5,5){$W^{\prime}_{\text{u}1}$}%
\rput(9,2.2){$W^{\prime}_{\text{u}2}$}%
\rput(12,2.2){$W^{\prime}_{\text{u}3}$}%
\rput[l](9.3,3.8){$R_{2},T_{2}$}%
\rput[l](12.3,3.8){$R_{2},T_{2}$}%
\end{pspicture}%
%
\caption{Noisy one-port with resistors at different temperatures \cite{Zinke, Schiek}}%
\label{noisyoneport}%
\end{figure}
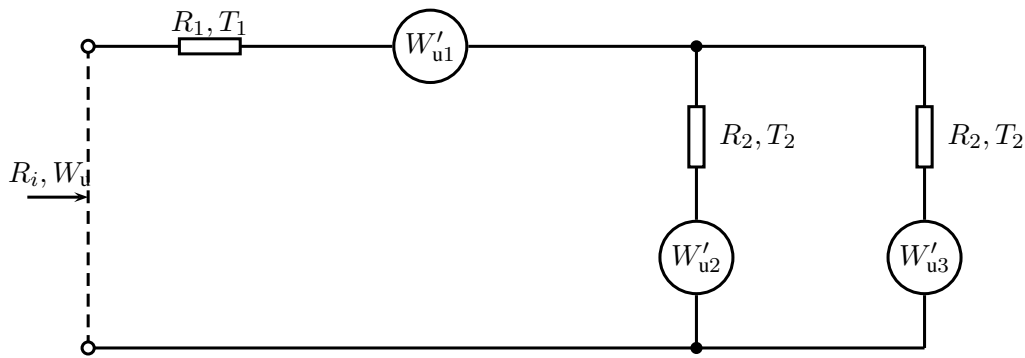

\begin{figure}[htbp]%
\centering
\begin{pspicture}(0,0)(15,4.5)%
\psset{linewidth=0.04}%
\psline(1,3)(8,3)
\pscircle[fillstyle=solid](2.5,3){0.5}
\pscircle[fillstyle=solid](4.5,3){0.5}
\pscircle[fillstyle=solid](6.5,3){0.5}
\psbrace(2,2.4)(7,2.4){}
\rput(4.5,1.5){$W_{\text{u}}$}%
\psline(1,1)(8,1)
\pscircle[fillstyle=solid](1,1){0.1}
\pscircle[fillstyle=solid](1,3){0.1}
\psline(8,1)(8,1.6)
\pspolygon(7.9,1.6)(8.1,1.6)(8.1,2.4)(7.9,2.4)
\psline(8,2.4)(8,3)
\rput[l](8.2,2){$R_{i}$}
\psline(9.4,1.9)(9.8,1.9)%
\psline(9.4,2.1)(9.8,2.1)%
\psline(11,1)(14,1)%
\psline(11,3)(14,3)%
\pscircle[fillstyle=solid](11,1){0.1}
\pscircle[fillstyle=solid](11,3){0.1}
\rput(6,0){%
\psline(8,1)(8,1.6)
\pspolygon(7.9,1.6)(8.1,1.6)(8.1,2.4)(7.9,2.4)
\psline(8,2.4)(8,3)
\rput[l](8.2,2){$R_{i}$}%
}%
\pscircle[fillstyle=solid](12.5,3){0.5}%
\rput(12.5,3){$W_{\text{u}}$}%
\rput(2.5,3){$W_{\text{u}1}$}%
\rput(4.5,3){$W_{\text{u}2}$}%
\rput(6.5,3){$W_{\text{u}i}$}%
\end{pspicture}%
%
\caption{Equivalent sources for the circuit of Fig.\,\ref{noisyoneport}.}%
\label{equivsources}%
\end{figure}
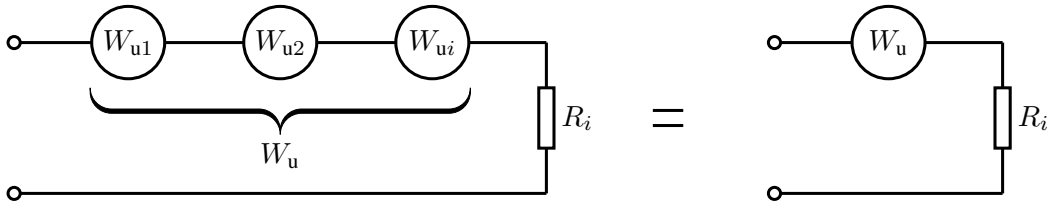

We define a spectral density function \cite{Schiek}
\begin{eqnarray}
\nonumber W_{\text{u}}(f) &=& 4kTR \\
W_{\text{i}}(f) &=& 4kTG \\
\nonumber \overline{u^{2}(t)} &=& \int_{f_{1}}^{f_{2}}W_{\text{u}}(f) \text{d}f\hspace{0.2cm}.
\label{specdensfunc}
\end{eqnarray}
A noisy resistor may be composed of many elements (resistive network). In general, it is made from many carbon grains which have homogeneous temperatures. But if we consider a network of resistors with different temperatures and hence with an inhomogeneous temperature distribution (Fig.\,\ref{noisyoneport})
%
%
the spectral density function changes to
\begin{eqnarray}
W_{\text{u}} = \sum_{j}W_{\text{u}j} = 4kT_{\text{n}}R_{\text{i}}\hspace{0.2cm}, \\
\label{multires1}
T_{\text{n}} = \sum_{j} \beta_{j}T_{j}\hspace{0.2cm},
\label{multires2}
\end{eqnarray}
where $W_{\text{u}j}$ are the noise sources (Fig.\,\ref{equivsources}), $T_{\text{n}}$ is the total noise temperature, $R_{\text{i}}$ the total input impedance, and $\beta_{j}$ are coefficients indicating the fractional part of the input power dissipated in the resistor $R_{j}$. It is assumed that the $W_{\text{u}j}$ are uncorrelated for reasons of simplicity.
%
%

The relative contribution ($\beta_{j}$) of a lossy element to the total noise temperature is equal to the relative dissipated power multiplied by its temperature:
\begin{equation}
T_{\text{n}} = \beta_{1}T_{1} + \beta_{2}T_{2} + \beta_{3}T_{3} + \cdots 
\label{relcont}
\end{equation}
A nice example is the noise temperature of a satellite receiver, which is nothing else than a directional antenna. The noise temperature of free space amounts roughly to 3\,K. The losses in the atmosphere, which is an air layer of 10 to 20\,km length, cause a noise temperature at the antenna output of about 10 to 50\,K. This is well below room temperature of 290\,K. 

So far only pure resistors have been considered. Looking at complex impedances, it can be seen that losses from dissipation occur in $Re$(Z) only. The available noise power is independent of the magnitude of Re($Z$) with Re($Z$) > 0. For Figs.\,\ref{noisyoneport} and \ref{equivsources}, Eq.\,(\ref{multires1}) still applies, except that $R_{\text{i}}$ is replaced by $Re$(Z$_{\text{i}}$). However, it must be remembered that in complex impedance networks the spectral power density $W_{\text{u}}$ becomes frequency dependent \cite{Zinke}.

The rules mentioned above apply to passive structures. A forward-biased Schottky diode (external power supply) has a noise temperature of about $T_{0}$/2 + 10\%. A biased Schottky diode is not in thermodynamic equilibrium and only half of the carriers contribute to the noise \cite{Schiek}. But it represents a real 50\,$\Omega$ resistor when properly forward biased. For transistors and in particular field-effect transistors (FETs), the physical mechanisms are somewhat more complicated. Noise temperatures of 50\,K have been observed on a FET transistor at 290\,K physical temperature.
\section{Noise-figure measurement with the spectrum analyser}
Consider an ideal amplifier (noiseless) terminated at its input (and output) with a load at 290\,K with an available power gain ($G_{\text{a}}$). We measure at the output \cite{Yip, HP}:
\begin{equation}
P_{\text{a}} = kT_{0}\Delta fG_{\text{a}}\hspace{0.2cm}.
\label{output}
\end{equation}

For $T_{0}$ = 290\,K (or often 300\,K) we obtain $kT_{0}$ = $-$174\,dBm/Hz ($-$dBm = decibel below 1\,mW). At the input we have for some signal S$_{i}$ a certain signal/noise ratio S$_{i}/\text{N}_{i}$ and at the output S$_{0}/\text{N}_{0}$. For an ideal (= noiseless) amplifier S$_{i}/\text{N}_{i}$ is equal to S$_{0}/\text{N}_{0}$, i.\,e., the signal and noise levels are both shifted by the same amount. This gives the definition of the noise figure $F$:
\begin{equation}
F = \frac{\text{S}_{i}/\text{N}_{i}}{\text{S}_{0}/\text{N}_{0}}\hspace{0.2cm}.
\label{noisefig}
\end{equation}

The ideal amplifier has $F$ = 1 or $F$ = 0\,dB and the noise temperature of this amplifier is 0\,K. The real amplifier adds some noise which leads to a decrease in $\text{S}_{0}/\text{N}_{0}$ due to the noise added (= N$_{\text{ad}}$):
\begin{equation}
F = \frac{\text{N}_{\text{ad}} + kT_{0}\Delta fG_{\text{a}}}{kT_{0}\Delta fG_{\text{a}}}\hspace{0.2cm}.
\label{adnoise}
\end{equation}

For a linear system the minimum noise figure amounts to $F_{\text{min}}$ = 1 or 0\,dB. However, for nonlinear systems one may define noise figures $F$ < 1. Now assume a source with variable noise temperature connected to the input and measure the
linear relation between amplifier output power and input termination noise temperature ($T_{\text{s}}$ = $T_{\text{source}}$).

\begin{figure}[htbp]%
\centering
\begin{pspicture}(0,0)(14,10)%
\psset{linewidth=0.04}%
\newrgbcolor{myred}{0.8 0.2 0}%
\newrgbcolor{olive}{0.6 0.6 0}%
\rput(0.5,0.5){%
\psline[arrows=->](-0.4,0.2)(12,0.2)%
\psline[arrows=->](1,0.2)(1,8)%
\psline[linestyle=dashed,linecolor=olive](1,0.2)(12,7.2)%
\pscircle[fillstyle=solid](1,0.2){0.1}%
\psline(-0.33333,0)(12,7.8)%
\rput[r](0.7,7.8){P$_{\text{OUT}}$}%
\rput[l](12.2,0.2){T$_{S}$ [K]}%
\rput(6.5,-0.2){Source temperature}%
}%
\rput(6.2,8.7){\small NOISE FREE}%
\rput(4.6,8.6){\small +}%
\rput(4.6,8.2){\small +}%
\rput[r](12,8.2){Slope $\approx k \text{G}_{\text{a}}\Delta$f}%
\rput[l](9,5){\olive Noise free}%
\rput[l](2.9,1.2){\myred T$_{0}$ = 300\,K}%
\rput[l](7.6,1.2){\myred T$_{\text{hot}}$ = 12000\,K \black (Drawing not to scale)}%
\psline[linecolor=myred](1.5,2)(2.6,2)%
\psline[linecolor=myred](2.6,2)(2.6,0.7)%
\pscircle[linecolor=myred,fillstyle=solid](2.6,2){0.1}%
\psline[linecolor=myred](1.5,5)(7.3,5)%
\psline[linecolor=myred](7.3,5)(7.3,0.7)%
\pscircle[linecolor=myred,fillstyle=solid](7.3,5){0.1}%
\multirput(0,0)(0.8,0){2}{%
\psline[linewidth=0.1](2.2,6.4)(2.6,6.4)%
\psline(2.4,6.4)(2.4,6.8)%
\pspolygon(2.3,6.8)(2.5,6.8)(2.5,7.6)(2.3,7.6)%
}%
\psline(2.4,7.6)(2.4,8.6)%
\psline(3.2,7.6)(3.2,8.2)%
\psline(2.4,8.6)(4.4,8.6)%
\psline(3.2,8.2)(4.4,8.2)%
\pspolygon(4.4,8.9)(4.4,7.9)(5.2,8.4)
\psline(5.2,8.4)(7.2,8.4)%
\pspolygon[linestyle=dashed](2.8,6.2)(7.4,6.2)(7.4,9.1)(2.8,9.1)%
\end{pspicture}%
%
\caption{Relation between source noise temperature $T_{\text{s}}$ and output power $P_{\text{out}}$ for an ideal (noise free) and a real amplifier \cite{Yip, HP}}%
\label{tempoutput}%
\end{figure}
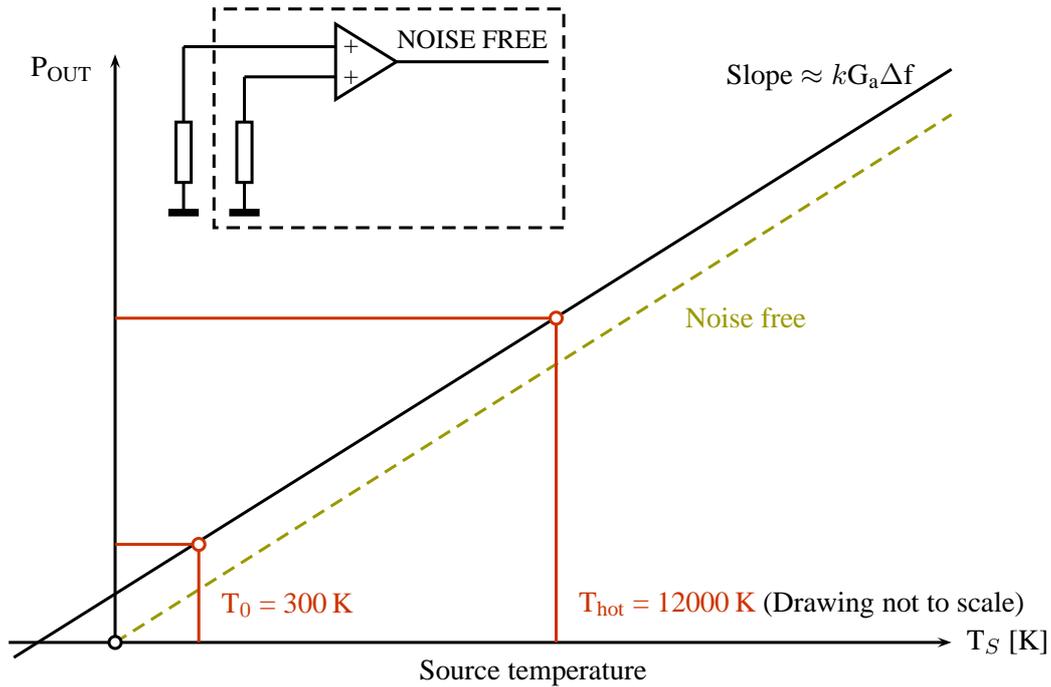

In a similar way a factor $Y$ can be defined:
\begin{eqnarray}
\nonumber Y = \frac{T_{\text{e}} + T_{\text{H}}}{T_{\text{e}} + T_{\text{C}}} \\
\nonumber T_{\text{e}} = \frac{T_{\text{H}} - Y T_{\text{C}}}{Y - 1} \\
F = \frac{\left[T_{(\text{H}}/290) - 1\right] - Y\left[(T_{\text{C}}/290) - 1\right]}{Y - 1}\hspace{0.2cm},
\label{yfactor}
\end{eqnarray}
where $T_{\text{e}}$ is the effective input noise temperature (see Fig.\,\ref{tempoutput}) and $T_{\text{H}}$ and $T_{\text{C}}$ are the noise temperatures of a hot or cold input termination.
To find the two points on the straight line of Fig.\,\ref{tempoutput} one may switch between two input terminations at 373\,K (100$^{\circ}$\,C) and 77\,K. For precise reading of RF power, calibrated piston attenuators in the IF path (Intermediate Frequency Superheterodyne Receiver) are in use. This is the hot/cold method. The difference between the $Y$-factor and the hot/cold method is that for the latter the input of the amplifier becomes physically connected to resistors at different temperatures (77, 373\,K). For the $Y$-factor, the noise temperature of the input termination is varied by electronic means between 300\,K and 12\,000\,K (physical temperature always around 300\,K).

As a variant of the 3-dB method with a controllable noise source, the excess noise temperature definition ($T_{\text{ex}} = T_{\text{H}} - T_{\text{C}}$) is often applied. A switchable 3-dB attenuator at the output of the amplifier just cancels the increase in noise power from $T_{\text{H}} - T_{\text{C}}$. Thus the influence of nonlinearities of the power meter is eliminated. To measure the noise of one port one may also use a calibrated spectrum analyser. However, spectrum analysers have high noise figures (20--40\,dB) and the use of a low-noise preamplifier is recommended. This `total power radiometer' \cite{Evans} is not very sensitive but often sufficient, e.\,g., for observation of the Schottky noise of a charged particle beam. Note that the spectrum analyser may also be used for two-port noise figure measurements. An improvement of this `total power radiometer' is the `Dicke Radiometer' \cite{Evans}. It uses a 1\,kHz switch between the unknown one port and a controllable reference source. The reference source is made equal to the unknown via a feedback loop, and one obtains a resolution of about 0.2\,K. Unfortunately, switch spikes sometimes appear. Nowadays, switch-free correlation radiometers with the same performance are available \cite{Schiek2}.

The noise figure of a cascade of amplifiers is \cite{Zinke, Schiek, Yip, HP, Schiek2}
\begin{equation}
F_{\text{total}} = F_{1} + \frac{F_{2} - 1}{G_{\text{a}1}} + \frac{F_{3} - 1}{G_{\text{a}1}G_{\text{a}2}} + \cdots
\label{cascade}
\end{equation}
As can be seen from Eq.\,(\ref{cascade}) the first amplifier in a cascade has a very important effect on the total noise figure, provided $G_{\text{a}1}$ is not too small and F$_{2}$ is not too large. In order to select the best amplifier from a number of different units to be cascaded, one can use the noise measure $M$:
\begin{equation}
M = \frac{F - 1}{1 - (1/G_{\text{a}})}\hspace{0.2cm}.
\label{noisemeasure}
\end{equation}
The amplifier with the smallest M has to be the first in the cascade \cite{HP}.
\end{document}